\documentclass[preprint, Astrosymb]{aastex631}

\usepackage{amsmath}

\newcommand{\pypeit}{\texttt{PypeIt} }

\shorttitle{Carbon-to-oxygen in brown dwarfs}
\shortauthors{Phillips et al.}
\graphicspath{{./}{figures/}}
\begin{document}

\title{The Carbon-to-Oxygen Ratio in Cool Brown Dwarfs and Giant Exoplanets. \\ I. The Benchmark Late-T dwarfs GJ 570D, HD 3651B and Ross 458C}

\author[0000-0001-6041-7092]{Mark W. Phillips}
\affiliation{Institute for Astronomy, University of Hawaii at Manoa, Honolulu, HI 96822, USA}
\correspondingauthor{Mark W. Phillips}
\email{mark.phillips@hawaii.edu}
%\affiliation{Astrophysics Group, University of Exeter, EX4 4QL, Exeter, UK}

\author[0000-0003-2232-7664]{Michael C. Liu}
\affiliation{Institute for Astronomy, University of Hawaii at Manoa, Honolulu, HI 96822, USA}

\author[0000-0002-3726-4881]{Zhoujian Zhang} \thanks{NASA Sagan Fellow}
\affiliation{Department of Astronomy \& Astrophysics, University of California, Santa Cruz, CA 95064, USA}

%% Note that the \and command from previous versions of AASTeX is now
%% depreciated in this version as it is no longer necessary. AASTeX 
%% automatically takes care of all commas and "and"s between authors names.

%% AASTeX 6.31 has the new \collaboration and \nocollaboration commands to
%% provide the collaboration status of a group of authors. These commands 
%% can be used either before or after the list of corresponding authors. The
%% argument for \collaboration is the collaboration identifier. Authors are
%% encouraged to surround collaboration identifiers with ()s. The 
%% \nocollaboration command takes no argument and exists to indicate that
%% the nearby authors are not part of surrounding collaborations.

%% Mark off the abstract in the ``abstract'' environment. 
\begin{abstract}

Measurements of the C/O ratio in brown dwarfs are lacking, in part due to past models adopting solar C/O only. We have expanded the \texttt{ATMO 2020} atmosphere model grid to include non-solar metallicities and C/O ratios in the T dwarf regime. We change the C/O ratio by altering either the carbon or oxygen elemental abundances, and we find that non-solar abundances of these elements can be distinguished based on the shapes of the $H$- and $K$- bands. We compare these new models with medium-resolution ($R\approx1700$), near-infrared ($0.8-2.4\,\mu$m) Gemini/GNIRS spectra of three benchmark late-T dwarfs, GJ 570D, HD 3651B, and Ross 458C. We find solar C/O ratios and best-fitting parameters ($T_\mathrm{eff}$, $\log(g)$, $Z$) broadly consistent with other analyses in the literature based on low-resolution ($R\sim100$) data. The model-data discrepancies in the near-infrared spectra are consistent across all three objects. These discrepancies are alleviated when fitting the Y, J, H and K bands individually, but the resulting best-fit parameters are inconsistent and disagree with the results from the full-spectrum. By examining the model atmosphere properties we find this is due to the interplay of gravity and metallicity on $\mathrm{H_2-H_2}$ collisionally induced absorption.  We therefore conclude that there are no significant issues with the molecular opacity tables used in the models at this spectral resolution. Instead, deficiencies are more likely to lie in the model assumptions regarding the thermal structures. Finally, we find a discrepancy between the GNIRS, SpeX, and other near-infrared spectra in the literature of Ross 458C, indicating potential spectroscopic variability. 

\end{abstract}

%% Keywords should appear after the \end{abstract} command. 
%% The AAS Journals now uses Unified Astronomy Thesaurus concepts:
%% https://astrothesaurus.org
%% You will be asked to selected these concepts during the submission process
%% but this old "keyword" functionality is maintained in case authors want
%% to include these concepts in their preprints.
%\keywords{Classical Novae (251) --- Ultraviolet astronomy(1736) --- History of astronomy(1868) --- Interdisciplinary astronomy(804)}

%% From the front matter, we move on to the body of the paper.
%% Sections are demarcated by \section and \subsection, respectively.
%% Observe the use of the LaTeX \label
%% command after the \subsection to give a symbolic KEY to the
%% subsection for cross-referencing in a \ref command.
%% You can use LaTeX's \ref and \label commands to keep track of
%% cross-references to sections, equations, tables, and figures.
%% That way, if you change the order of any elements, LaTeX will
%% automatically renumber them.
%%
%% We recommend that authors also use the natbib \citep
%% and \citet commands to identify citations.  The citations are
%% tied to the reference list via symbolic KEYs. The KEY corresponds
%% to the KEY in the \bibitem in the reference list below. 

\section{Introduction} \label{sec:intro}

Spectroscopy of cool brown dwarfs and giant exoplanets allows us to study the physics and chemistry of their atmospheres. Measuring spectroscopic signatures of composition has long been cited as a method of inferring the formation mechanism of planetary objects \citep[e.g.][]{Gaidos_2000, Lodders_2004, Kuchner_Seager_2005}. Of particular importance is the C/O ratio, which in exoplanets can depart from that of their host stars due to their formation and evolution within a protoplanetary disk, thus providing an observational link between present-day atmospheric composition and formation pathway. Compared to the solar value of $\mathrm{C/O}\approx0.55$ \citep{Asplund_2009}, the C/O ratio of exoplanets can become elevated as high as $\mathrm{C/O}\approx1$ when forming beyond the snowlines of carbon- and oxygen-rich ices \citep{Oberg_2011, Madhusudhan_2011, Molliere_2022}. Stellar abundances of C and O establish the initial conditions for disk chemistry in planet-formation models \citep{Bond_2010}, thus motivating studies of the stellar C/O ratio in the solar neighborhood. In solar-type and low-mass stars (including planet-hosting stars), the C/O ratio has been found to vary between $\approx0.3-0.8$ \citep{Nissen_2014, Nakajima_Sorahana_2016, Brewer_Fischer_2016, Suarez-Andres_2018}. Such C/O determinations have been obtained from high resolution spectroscopy (see \citet{Nissen_Gustafsson_2018} for a review), relying on weak individual absorption lines of C and O. 

In contrast, the dominant absorbers in cool brown dwarf atmospheres are water, methane and carbon monoxide, which carve out broad absorption bands across the infrared and account for almost all of the atmospheric C and O. Brown dwarfs therefore offer a unique diagnostic of the C/O ratio in the solar neighbourhood, arising from a different atmospheric chemistry than higher mass stars \citep{Fortney_2012}. Furthermore, wide-separation ($>100\,$au) brown dwarf companions to solar-type stars provide crucial benchmarks for such investigations thanks to the age and composition constraints derived from their host stars \citep[e.g.][]{Crepp_2018, Rickman_2020, Zhang_2020}. Comparing the measured C/O ratio of a brown dwarf companion to that of the host star could shed light on the companion's formation pathway, e.g. planet-like formation in a disk or star-like formation from cloud fragmentation. However, measurements of the C/O ratio in brown dwarfs, both companions and free-floating objects, are lacking in part due to past models encoding solar C/O into the model assumptions. 

Over the last few years, new grids of coupled atmosphere and evolution models have been developed \citep{Phillips_2020, Marley_2021}. These models include the latest molecular opacities and treatments for atmospheric chemistry, benefitting from a wide range of developments since the last major set of models over a decade ago. The improvements to the chemistry are particularly important for accurately determining the C/O ratio and non-solar abundances. First, the addition of non-equilibrium chemistry that is self-consistent with the pressure-temperature profile is vital \citep{Drummond_2016}, as mixing processes can drive the abundances of carbon- and oxygen bearing molecules away from their equilibrium values. Second, the addition of rainout chemistry, which models the depletion of elements due to the settling/sinking of condensates in an atmosphere \citep{Burrows_Sharp_1999}, is also important since the formation of condensates can deplete up to 20\% of the oxygen atoms from the observable atmosphere \citep{Lodders_2010}, impacting the observed C/O ratio. This new generation of atmosphere models allows us to more reliably investigate the spectroscopic signatures of non-solar chemical abundances in cool brown dwarfs.

The majority of brown dwarf spectroscopy has been conducted at low spectral resolution ($R<1000$) \citep{Leggett_2000, Sorahana_2014, Burgasser_2014, Schneider_2015}. Many observations have been taken with the SpeX spectrograph \citep{Rayner_2003} on the NASA Infrared Telescope Facility (IRTF), which provides $R\approx75-120$ near-infrared ($0.8-2.5\,\mu$m) spectra in its prism mode. These observations have led to large compilations such as the SpeX Prism Library \citep{Burgasser_2014}, which contains thousands of spectra of ultracool M, L, and T dwarfs. The $0.8-2.5\,\mu$m wavelength range provided by SpeX captures the majority of the spectral energy distribution of M, L and T-type ultracool dwarfs. Mid-late T dwarfs are perhaps the most promising objects to constrain C/O ratios, with the clouds and/or thermochemical instabilities that complicate the modeling and studies of earlier spectral types having sunk below the photosphere. The numerous SpeX prism spectra available in the literature have led to many T dwarf model-data comparisons, using fully consistent one-dimensional radiative-convective atmosphere models \citep{Oreshenko_2020, Zhang_2021_a, Zhang_2021_b} and data-driven atmospheric retrieval frameworks \citep{Line_2015, Line_2017, Zaleksy_2019, Kitzmann_2020, Zalesky_2022, Whiteford_2023} to interpret the atmospheric properties.  However, at low spectral resolution, molecular absorption lines are not resolved, so high resolutions are required to test the opacities and line lists used in the atmosphere models.  A small sample of T dwarfs have been observed at medium-resolution with the Folded-Port Infrared Echelette (FIRE) spectrograph ($R\approx6000$; $0.85-2.5\,\mu$m) at the Magellan telescopes \citep{Burgasser_2011, Bochanski_2011, Hood_2023} and the Near-Infrared Integral Field spectrograph (NIFS) ($R\approx5000$; $1.0-2.4\,\mu$m) at the Gemini North telescope \citep{Canty_2015}. T dwarfs have been observed at high resolution with IGRINS ($R\approx45,000$; $1.46-2.48\,\mu$m; \citealp{Tannock_2022}) and with NIRSPEC ($R\approx35,000$; $2.29-2.49\,\mu$m; \citealp{Xuan_2022}) and ($R\approx20000$; $1.165-323\,\mu$m; \citealp{McLean_2007}). However, these studies typically focused on a single object and/or had a limited wavelength range. There is thus a need for higher-resolution spectra than provided by SpeX near-infrared for a systematic sample of cool brown dwarfs.

To address this, we are conducting a spectroscopic survey of mid/late T dwarfs obtaining medium-resolution ($R\approx1700$), high signal-to-noise near-infrared spectra ($0.8-2.4\,\mu$m) using the Gemini Near-Infrared Spectrograph (GNIRS). In this work, we present GNIRS spectra for the benchmark brown dwarfs GJ 570D, HD 3651B and Ross 458C, which are all wide-separation companions to higher mass stars. To analyse and interpret these spectra, we have expanded the \texttt{ATMO} 2020 model atmosphere grid \citep{Phillips_2020} to include non-solar abundances, through changing the bulk metallicity and C/O ratio of the model atmospheres in the mid/late T dwarf regime ($T_\mathrm{eff}=700-900\,$K). The paper is organised as follows. In Section \ref{sec:spectroscopy} we outline the details of our GNIRS observations and our data reduction methods. In Section \ref{sec:atmo} we discuss the details of the \texttt{ATMO} model atmosphere grid. The results of fitting the \texttt{ATMO} models to the GNIRS spectra of GJ 570D, HD 3651B and Ross 458C are presented in Section \ref{sec:results}. We then analyse and discuss the discrepancies between the models and data in Section \ref{sec:discussion}, before summarizing our work in Section \ref{sec:conclusions}.

\section{Methods} \label{sec:methods}

\subsection{Gemini/GNIRS spectroscopy} \label{sec:spectroscopy}

We construct a sample of cool brown dwarfs from the UltracoolSheet \citep{Best_2020} which contains astrometry, photometry, spectral classifications and multiplicities for 3000+ ultracool dwarfs and imaged exoplanets from multiple sources and surveys \citep{Dupuy_Liu_2012, Dupuy_Kraus_2013, Liu_2016, Best_2018, Best_2021}. We select spectral types T6 and later, which correspond to $T_\mathrm{eff}\lesssim1000\,\mathrm{K}$ \citep{Filippazzo_2015}, since the clouds that complicate the modeling of earlier spectral types are believed to have sunk below the photosphere for late-T dwarfs. We then apply a cut in apparent magnitude of $J<16$ to select the brightest objects, relaxing this to $J<17$ for objects of interest such as benchmark companions and young moving group members. 

This paper focuses on the benchmark brown dwarfs GJ 570D, HD 3651B and Ross 458C, with the wider sample being the subject of future papers. We observed these objects at the Gemini-North 8.1-m telescope with the facility near-infrared spectrograph GNIRS \citep{Elias_2006}, using the short blue camera's cross-dispersed (SXD) mode with the $32\,$l/mm grating and $0.3"$ slit aligned to the parallactic angle, providing $R\approx1700$. We took four $189\,$s exposures of GJ 570D,  eight $280\,$s exposures of HD 3651B, and sixteen $326\,$s exposures of Ross 458C, dithered in an ABBA pattern along the slit. We observed the standard stars HIP 73867 ($V=8.6\,$mag),  HIP 6193 ($V=4.7\,$mag)  and HIP 68868 ($V=8.6\,$mag) for telluric correction of GJ 570D,  HD 3651B,   and Ross 458C, respectively, using the same instrument setup and dither pattern. The observations are summarized in Table \ref{tab:gnirs_observations}.

We reduce the Gemini/GNIRS spectra using the open-source Python package \pypeit \citep{Prochaska_2020_arxiv, Prochaska_2020_zenodo, Bochanski_2009, Bernstein_2015}. \pypeit is a semi-automated spectroscopic data reduction pipeline that can be applied to slit-imaging spectrographs and supports longslit, multislit and cross-dispersed echelle spectra. To reduce spectra, \pypeit requires calibration flat frames for order edge tracing and pixel efficiency correction, as well as arc frames for wavelength calibration. We use 6 nighttime flats taken with a quartz-halogen lamp for the flat frames. \pypeit determines the wavelength solution in the infrared from OH sky lines, which is preferable to using the images of the Ar lamp due to flexure. Therefore the science frames are used to perform the wavelength calibration and to trace the tilt in the wavelength solution across the slit. However, for short exposures ($<60\,$s) the OH sky lines are too weak, and thus for the telluric standard star \pypeit uses the tilt and wavelength solution from the nearest-in-time science image. The OH sky lines provide robust wavelength solutions, with the typical RMS for each order being $\le0.1\,$pixels ($6\times10^{-5}\,\mu$m). \pypeit achieves Poisson limited sky-subtraction \citep{Prochaska_2020_arxiv} and then performs an optimal extraction to generate 1D science spectra.

Flux calibration and telluric correction are performed using an infrared sensitivity function. The sensitivity function is generated from the observations of the standard star, by comparing the photon flux of the standard with its absolute, previously calibrated flux. In the case of the infrared observations conducted in this work, we use the Vega spectrum to generate the sensitivity function. The telluric model is similarly derived by fitting to a grid of model atmosphere spectra, with a range of atmospheric conditions specific to different observatories, including Mauna Kea. The sensitivity function can then be applied to the science observations which converts the measured photons per second per $\mathrm{\AA}$ to flux units ($\mathrm{erg\,s^{-1}\,\AA^{-1}\,cm^{-2}}$), followed by telluric correction. 

After extraction, \texttt{PypeIt} rebins the spectra onto a common logarithmic wavelength grid, which facilitates the coaddition of multiple exposures and the combination of overlapping echelle orders. Multiple science exposures of the same object are combined for each echelle order by taking their average, weighted by the square of the median $\mathrm{S/N}$ ratio. To do this, a scale factor which normalizes the spectra to a common intensity, is calculated from the frame with the largest signal. Finally, to combine the edges of overlapping echelle orders, \texttt{PypeIt} averages the data weighting by the inverse variance of each pixel. For more information on the algorithms used in the \texttt{PypeIt} data reduction pipeline, we refer the reader to \citet{Bochanski_2009}, \citet{Bernstein_2015}, and the \texttt{PypeIt} online documentation\footnote{\url{https://pypeit.readthedocs.io}}.

\begin{figure}[ht!]
\plotone{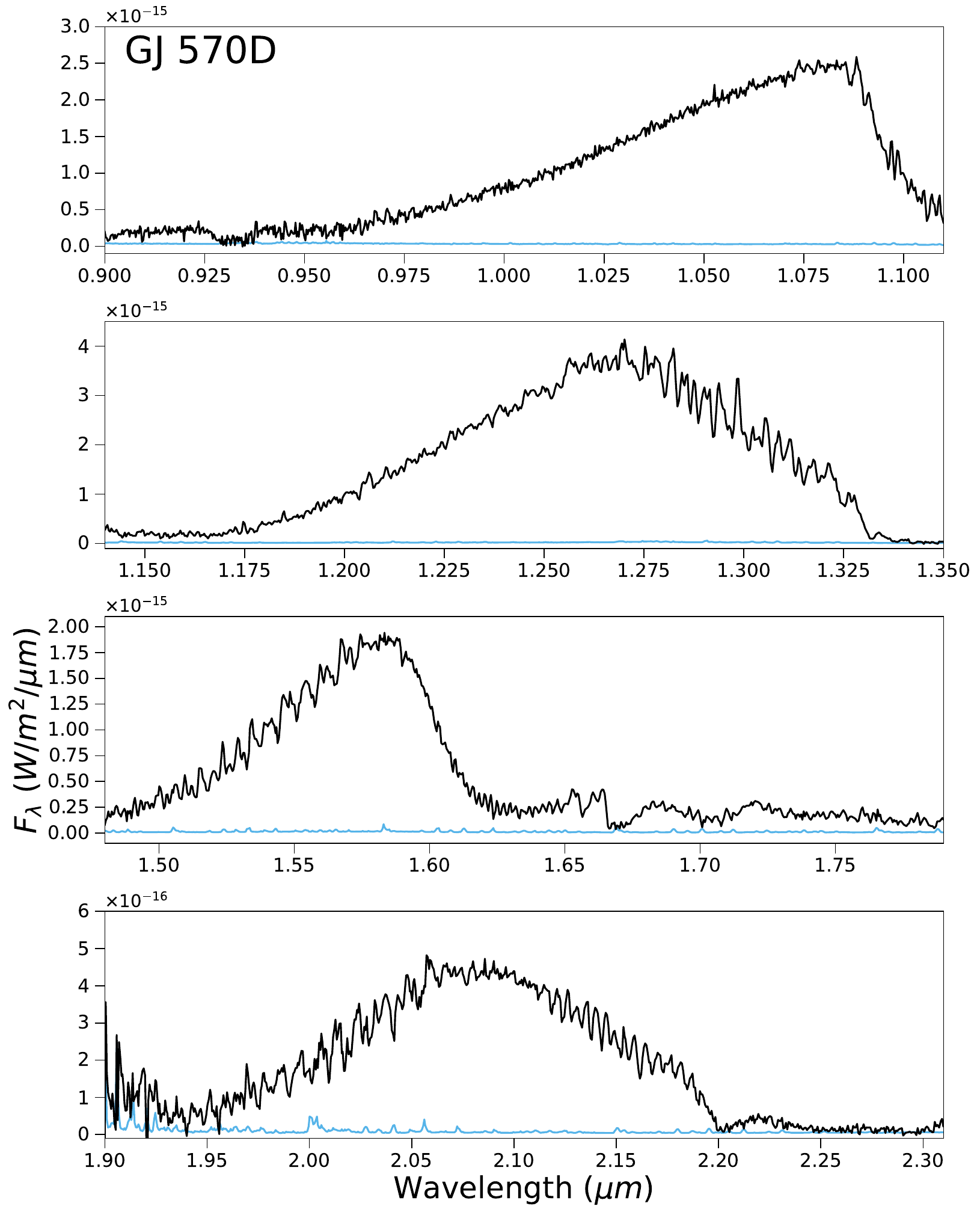}
\caption{Gemini North/GNIRS spectrum of GJ 570D (black). The uncertainty on the GNIRS spectrum is shown in light blue. \label{fig:GJ_570D}}
\end{figure}

\begin{figure}[ht!]
\plotone{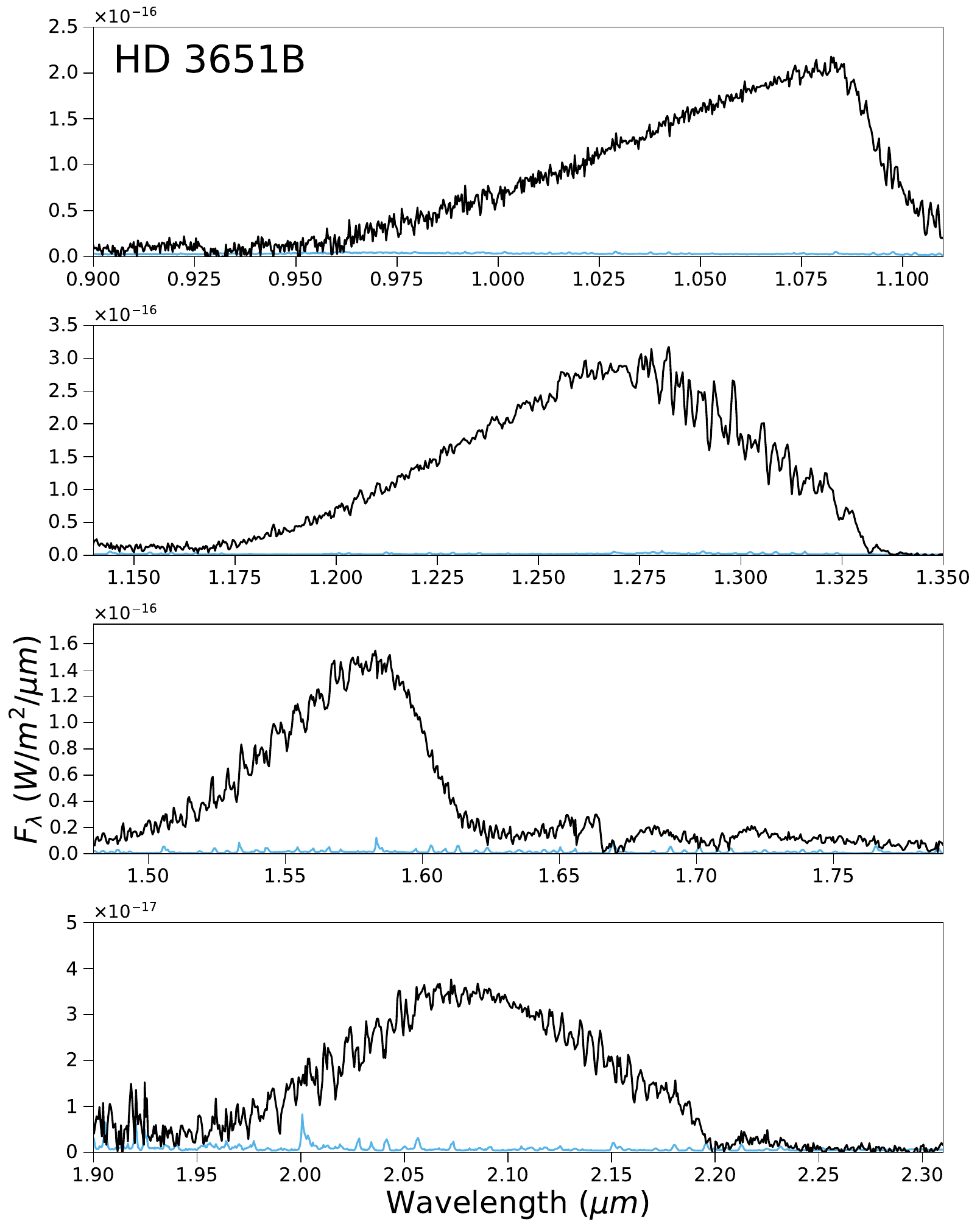}
\caption{ Gemini North/GNIRS spectrum of HD 3651B (black). The uncertainty on the GNIRS spectrum is shown in light blue. \label{fig:HD_3651B}}
\end{figure}

\begin{figure}[ht!]
\plotone{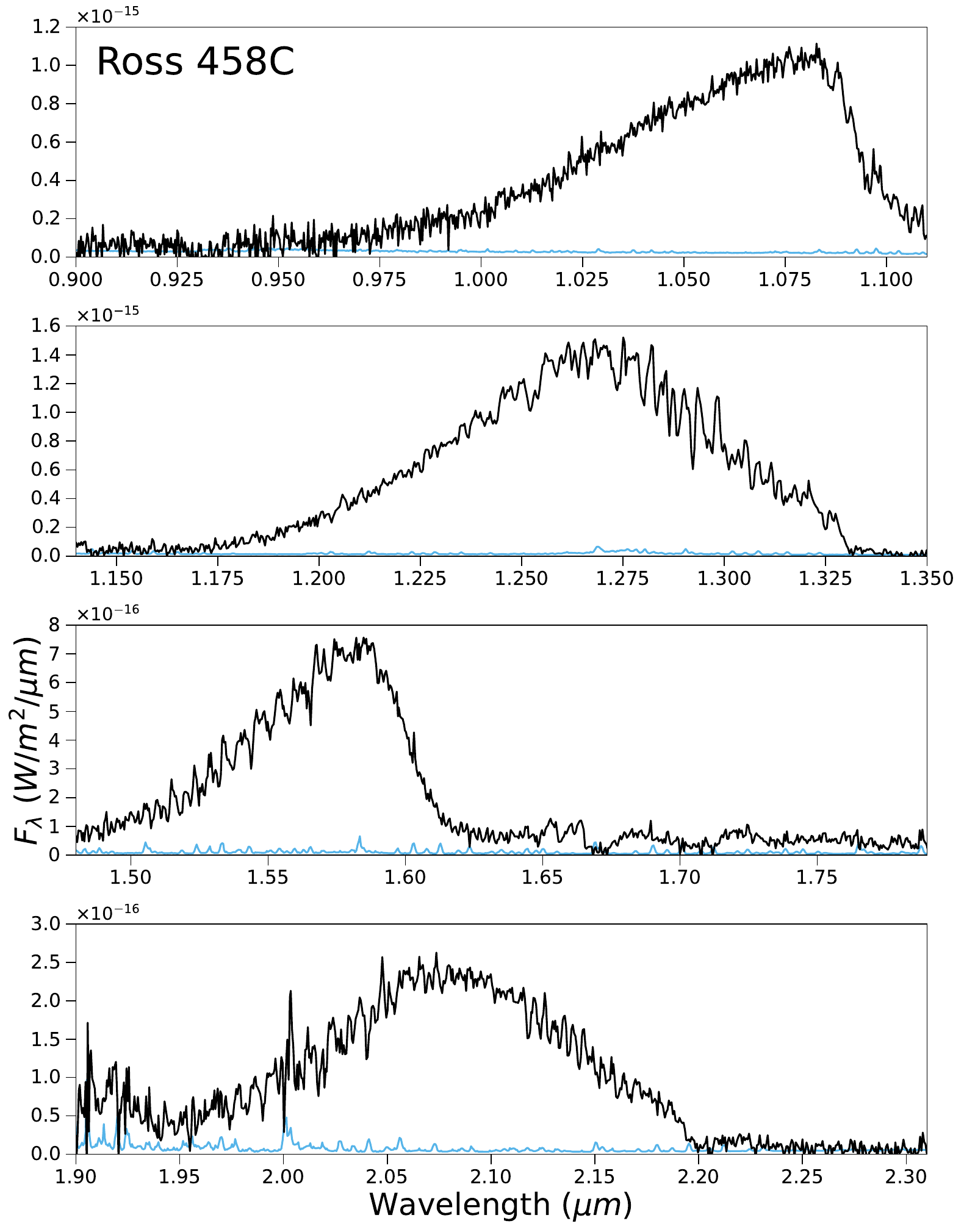}
\caption{Gemini North/GNIRS spectrum of Ross 458C (black). The uncertainty on the GNIRS spectrum is shown in light blue. \label{fig:ROSS_458C}}
\end{figure}

The final flux-calibrated, coadded, and telluric-corrected spectra of GJ 570D, HD 3651B  and Ross 458C are shown in Figures \ref{fig:GJ_570D}, \ref{fig:HD_3651B} and \ref{fig:ROSS_458C}, respectively. We flux-calibrate the spectra to the MKO $H$-band magnitude of all objects following \citet{Zhang_2021_a}. 

\subsection{Atmosphere model} \label{sec:atmo}

To analyse the GNIRS spectra we use the recently published \texttt{ATMO} 2020 atmosphere models \citep{Phillips_2020}. Briefly, the \texttt{ATMO} code computes the pressure-temperature structure of a one-dimensional atmosphere by solving for radiative-convective and hydrostatic equilibrium in each model level. We adopt the same model setup described in detail in \citet{Phillips_2020} and references therein, which includes descriptions of the opacity database, radiative transfer, and chemistry schemes used by the \texttt{ATMO} code. Notably, the \texttt{ATMO} models include the latest line shapes for the potassium resonance doublet \citep{Allard_2016} and self-consistent non-equilibrium chemistry due vertical mixing. The initial tranche of publically available \texttt{ATMO} 2020 models were generated at solar metallicity. We have developed new models with non-solar compositions, through changing the bulk metallicity and C/O ratio of the atmosphere\footnote{Models made publicly available on \url{http://opendata.erc-atmo.eu}}. The grid spans $T_\mathrm{eff}=700-900\,$K in steps of $100\,$K, $\log(g)=4.5-5.5\,$dex in steps of $0.5\,$dex, $Z=-0.5-0.5\,$dex in steps of $0.5\,$dex, and $\mathrm{C/O}=0.35, 0.55$(solar), and $1.0$. In total there are 162 models. The C/O ratio can be changed by altering either the carbon or oxygen elemental abundances. In this work, we vary either the carbon or oxygen abundance through scaling factors $X_C$ and $X_O$, respectively, while keeping the other elemental abundances constant. The bulk metallicity is changed by scaling all elements other than H and He by a constant factor. In our models we adopt an eddy diffusion coefficient of $\log(K_{zz})=4$, which is held constant throughout both the radiative and convective regions of the atmosphere. We note that the value of $K_{zz}$ is uncertain by multiple orders of magnitude, particularly in the radiative region \citep{Mukherjee_2022}. We also compute some models with $\log(K_{zz})=6$ and $\log(K_{zz})=8$ to examine this parameter's effect on the model spectra. However in the grid-fitting analyses presented in Section \ref{sec:results}, we use only the $\log(K_{zz})=4$ models. We choose a value of $\log(K_{zz})=4$ to be consistent with literature values that have been found to reasonably match the observed colors of late T and Y dwarfs \citep{Leggett_2017}.

We use these models to interpret the observed GNIRS spectra. To do this, we first degrade the model spectra to the GNIRS resolution of $R=1700$. This is done by convolving the model spectrum with a Gaussian kernel, and then binning the spectrum onto the same wavelength grid as the observations. We then bi-linearly interpolate between the model grid points, such that the model sub-grid steps are $\Delta T_\mathrm{eff}=20\,K$, $\Delta\log(g)=0.1\,$dex, $\Delta Z=0.1\,$dex, $\Delta X_O=0.1$ (similarly for $\Delta X_C$), between the minimum and maxiumum value of each parameter giving $34,607$ models in total. Each model in the grid is then scaled by the geometric dilution factor $\alpha=R^2/D^2$, where $R$ is the radius of the object and $D$ is the distance to the object. The dilution factor can be obtained by setting $\frac{\partial\chi^2}{\partial\alpha}=0$, where $\chi^2$ is the chi-squared statistic, defined as 

\begin{equation}
    \chi^2=\sum^N_{i=1}\left(\frac{O_i-\alpha M_i}{\sigma_i}\right)^2.
    \label{eq:chisq}
\end{equation}

\noindent Here $O_i$ and $M_i$ are the observed and model flux densities, $\sigma_i$ is the uncertainties on the observed flux densities, and $N$ is the number of data points. The maximum-likelihood value of $\alpha$ is then 

\begin{equation}
    \alpha = \frac{\sum^N_{i=1}\frac{O_i M_i}{\sigma_i^2}}{\sum^N_{i=1}\frac{M_i^2}{\sigma_i^2}}.
\end{equation}

\noindent The best-fit model is then found by computing the $\chi^2$ statistic for all the models and then choosing the minimum value. We report our results as the reduced $\chi^2$, defined as $\chi_\nu^2=(1/\nu)\chi^2$ where $\nu$ is the number of degrees-of-freedom. We note that the values are large compared to the standard expectation of a good model fit ($\chi_\nu^2\approx1$). This is a well-known issue in the literature, given that $\chi^2$ values are influenced not only by the measurement uncertainties, but also the systematic deficiencies in the model assumptions. 

We adopt uncertainties on the best-fit parameters of half the model grid spacing (before interpolation). This is approximately consistent with the findings of \citet{Zhang_2021_a}, who found uncertainties typically in the range of $1/3-1/2$ when using the Bayesian inference tool \texttt{Starfish} combined with the Sonora-Bobcat \citep{Marley_2021} models to conduct an analysis of low-resolution T dwarf spectra. Uncertainties of half the model grid spacing have been commonly adopted when using a traditional $\chi^2$ based model fitting analysis \citep{Cushing_2008, Zhang_2020}, since the models are known to have systematic errors and thus Equation \ref{eq:chisq} will not follow a true $\chi^2$ distribution.

\section{Results}
\label{sec:results}

First we discuss the results of our atmosphere modeling and explore the spectral signatures of non-solar metallicities and C/O ratios in Section \ref{sec:model_results}. We then present the results of fitting these atmosphere models to our GNIRS spectra of GJ 570D, HD 3651B and Ross 458C in Sections \ref{sec:results_570D}, \ref{sec:results_3651B} and \ref{sec:results_458C}, respectively.

\subsection{Non-solar ATMO models}
\label{sec:model_results}

\begin{figure}[ht!]
\begin{center} 
\includegraphics[scale=0.6]{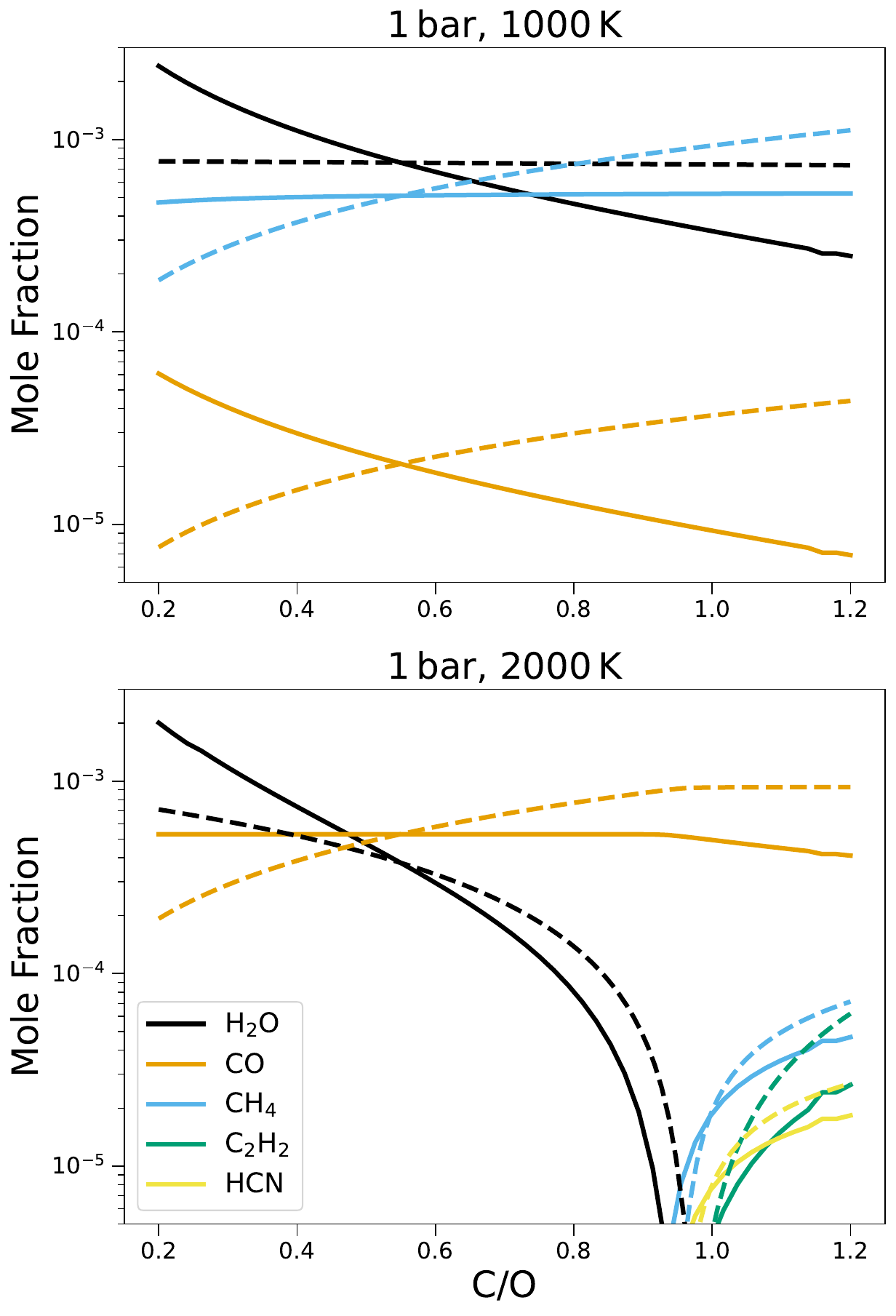}
\end{center}
\caption{Mole fractions of the dominant infrared opacity sources in cool brown dwarf atmospheres as a function of C/O ratio for two different temperatures.  The two panels approximately represent the temperatures at which the $Y$ and $J$ band ($\sim2000\,$K) and $H$ and $K$ band ($\sim1000\,K$) flux peaks originate. Solid and dashed lines correspond to varying the oxygen and carbon abundances respectively.\label{fig:CdO_chemistry}}
\end{figure}

Our new non-solar \texttt{ATMO} models allow us to explore the effect that changing the C/O ratio of the atmosphere has on the near-infrared spectrum. The dominant near-infrared absorbers $\mathrm{H_2O}$, $\mathrm{CH_4}$ and CO vary as a function of C/O ratio, as shown in Figure \ref{fig:CdO_chemistry}. $\mathrm{CH_4}$ and CO are the dominant carbon-bearing species at low ($<1300\,$K) and high ($>1300\,$K) temperature, respectively. Oxygen is predominantly contained within $\mathrm{H_2O}$ at low temperature, and in $\mathrm{H_2O}$ and CO at higher temperatures. It can be clearly seen that changing the C/O ratio, the abundances of these key molecules varies depending on whether the abundance of C or O is changed. 

When varying the oxygen abundance of the atmosphere, at lower temperatures $\mathrm{CH_4}$ is independent of C/O, and $\mathrm{H_2O}$ and CO both decrease with increasing C/O as the number of oxygen atoms decreases. At higher temperatures the abundance of CO is limited by the fixed carbon abundance, and is independent of C/O for $\mathrm{C/O}<1$. For $\mathrm{C/O}>1$, the decreasing oxygen abundance limits the CO abundance. The abundance of $\mathrm{H_2O}$ increases as C/O decreases due to the increasing number of oxygen atoms.  

When varying the carbon abundance of the atmosphere, at low temperatures $\mathrm{H_2O}$ is independent of C/O and $\mathrm{CH_4}$ increases with increasing C/O as the number of available carbon atoms increases. At higher temperatures, the abundance of CO increases with increasing C/O for $\mathrm{C/O}<1$ as the number of available carbon atoms increases, but become independent of C/O for $\mathrm{C/O}>1$ due to the fixed oxygen abundance. The $\mathrm{H_2O}$ abundance also varies when varying the carbon abundance at high temperatures, since it depends on the available oxygen atoms and thus the abundance of CO. For $\mathrm{C/O}>1$, when varying both carbon and oxygen abundances, trace species of $\mathrm{C_2H_2}$ and $\mathrm{HCN}$ become more abundant than $\mathrm{H_2O}$, due to the increasing number of free carbon atoms. These trace species have spectral features in the mid-infrared, at $\sim3$, $\sim7.5$ and $\sim14\,\mu$m. 

\begin{figure}[ht!]
\plotone{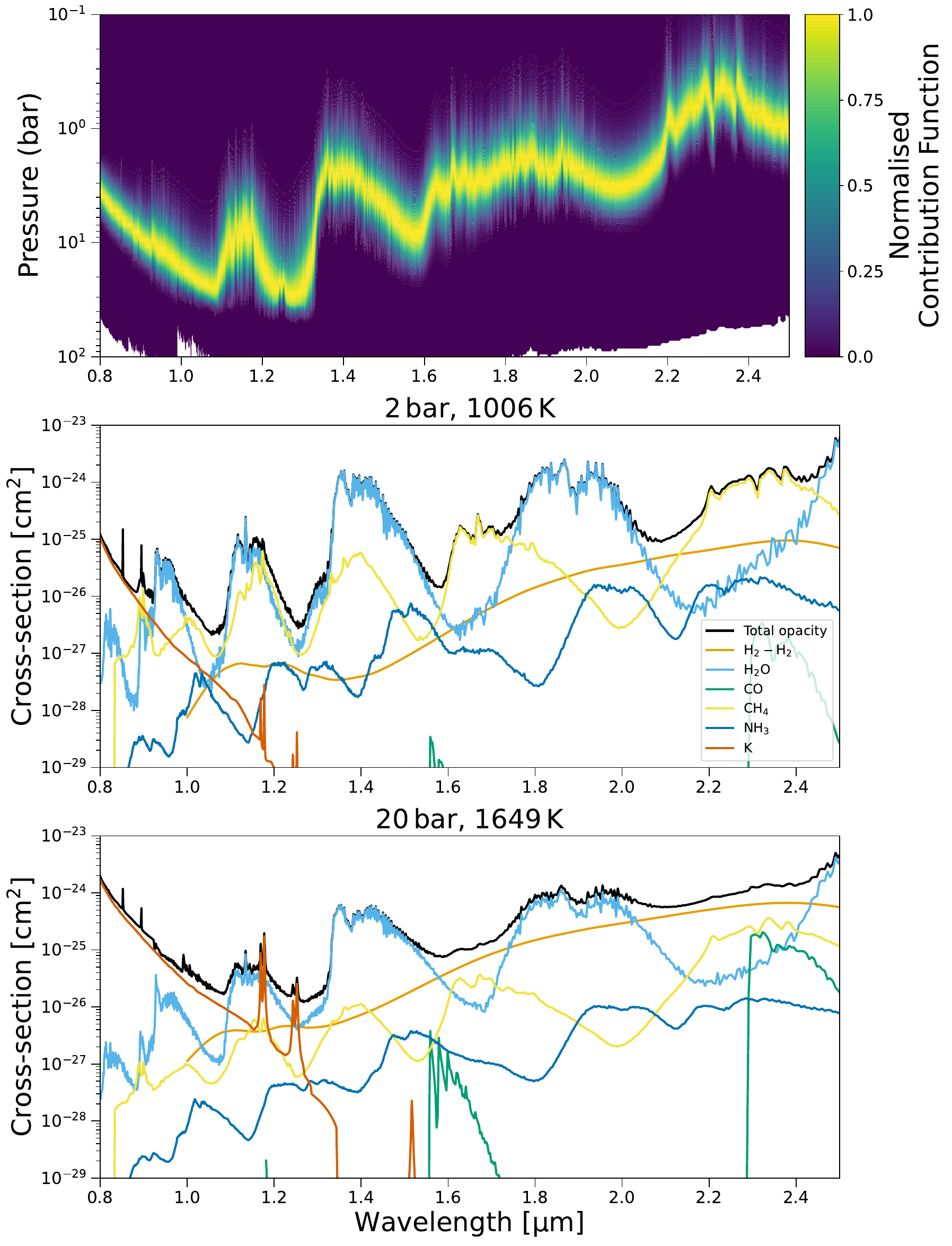}
\caption{Top: Contour plot of the normalised contribution function, illustrating the pressure levels of the model atmosphere contributing to the emission from the top of the atmosphere at a given wavelength. The model has $T_\mathrm{eff}=800\,$K, $\log(g)=4.5$, $Z=0.0$ and $\mathrm{C/O}=0.55$. Middle \& Bottom: Abundance weighted absorption cross-sections of the main atomic and molecular opacity sources in a typical T dwarf atmosphere, with the same parameters as the top panel. The two panels show the cross-sections at two different atmospheric levels. \label{fig:Tdwarf_opacities}}
\end{figure}

Figure \ref{fig:Tdwarf_opacities} shows the contribution function and abundance-weighted absorption cross-sections of the main atomic and molecular opacity sources in a typical T dwarf atmosphere from the \texttt{ATMO 2020} model grid ($T_\mathrm{eff}=800\,$K, $\log(g)=4.5$, assuming chemical equilibrium with solar elemental abundances). Combined with the variation of the molecular abundances with C/O ratio shown in Figure \ref{fig:CdO_chemistry}, we can interpret the impact the model grid parameters have on the observable near-infrared emission, shown in Figures \ref{fig:Yband_models} through \ref{fig:Kband_models}. Decreasing the effective temperature causes the absorption bands of $\mathrm{H_2O}$ and $\mathrm{CH_4}$ to become broader and deeper, impacting all four of the $Y$, $J$, $H$ and $K$ bands. Changing the surface gravity impacts the $K$ band, due to increased $\mathrm{H_2-H_2}$ collisionally induced absorption, which peaks at this wavelength range (see Figure \ref{fig:Tdwarf_opacities}).  We note that while $\mathrm{H_2-H_2}$ CIA peaks in the $K$-band, it can also be an important absorber in the $Y$, $J$ and $H$ bands. The $Y$ and $J$ band can probe pressures greater than $20\,$bar. At these pressures $\mathrm{H_2-H_2}$ CIA is the second most prominent absorber in the $Y$-band behind the broad red wings of the K I resonance doublet, which is often thought as being the sole opacity source in this wavelength range. In the $J$-band, $\mathrm{H_2-H_2}$ CIA is the second most important opacity source behind $\mathrm{H_2O}$ at high pressures. While the $H$-band does not probe as deep into the atmosphere as the $Y$ or $J$ bands, $\mathrm{H_2-H_2}$ CIA can be the most dominant absorber in the $H$-band at high pressures. $\mathrm{H_2-H_2}$ CIA therefore plays a role in shaping T-dwarf spectra across the near-infrared, not just in the $K$ band. It is particularly important in high-gravity atmospheres in which the peak of the contribution function shifts to higher pressures.  

 Changes in the metallicity of the atmosphere also impact the $\mathrm{H_2-H_2}$ collisionally induced absorption. Decreasing the metallicity leads to a relative increase in the abundance of H and He atoms, and a decrease in the abundances of important near-infrared molecular absorbers. This leads to a less opaque atmosphere, shifting the photosphere to higher pressures, and enhancing the $\mathrm{H_2-H_2}$ collisionally induced absorption. This effect primarily impacts the $K$ band where the $\mathrm{H_2-H_2}$ collisionally induced absorption peaks, but also the peak fluxes in all the bands. We note that the degeneracy between metallicity and surface gravity for T dwarf atmospheres has been quantified by \citet{Zhang_2021_a}. 

Similarly to metallicity, decreasing the oxygen abundance of the atmosphere decreases the abundance of $\mathrm{H_2O}$, increasing the relative abundance of H atoms and enhancing the $\mathrm{H_2-H_2}$ collisionally induced absorption in the $Y$ and $K$ band. Additionally, changing the oxygen abundance also manifests as changes in flux in the red side of the $H$ band as seen in Figure \ref{fig:Hband_models}, due to the increasing $\mathrm{CH_4}$ abundance as the oxygen abundance is decreased (see Figure \ref{fig:CdO_chemistry}, top panel). Altering the carbon abundance impacts either side of the $H$ band and also the blue side of the $K$ band. The methane abundance increases when increasing the carbon abundance leading to differences in the red side of the $H$ band. The $\mathrm{H_2O}$ abundance also decreases with increasing carbon abundance leading to differences in the blue side of the $H$ and $K$ bands. While these models indicate that non-solar carbon and oxygen abundances could be distinguished from the shapes of the $H$ and $K$ band, we found in our model fitting analysis that there are negligible differences in the spectra of the best-fitting models to GJ 570D when changing the C/O ratio through either the carbon or oxygen abundances (Section \ref{sec:results_570D}).

\subsection{GJ 570D}
\label{sec:results_570D}

\begin{figure}[ht!]
\plotone{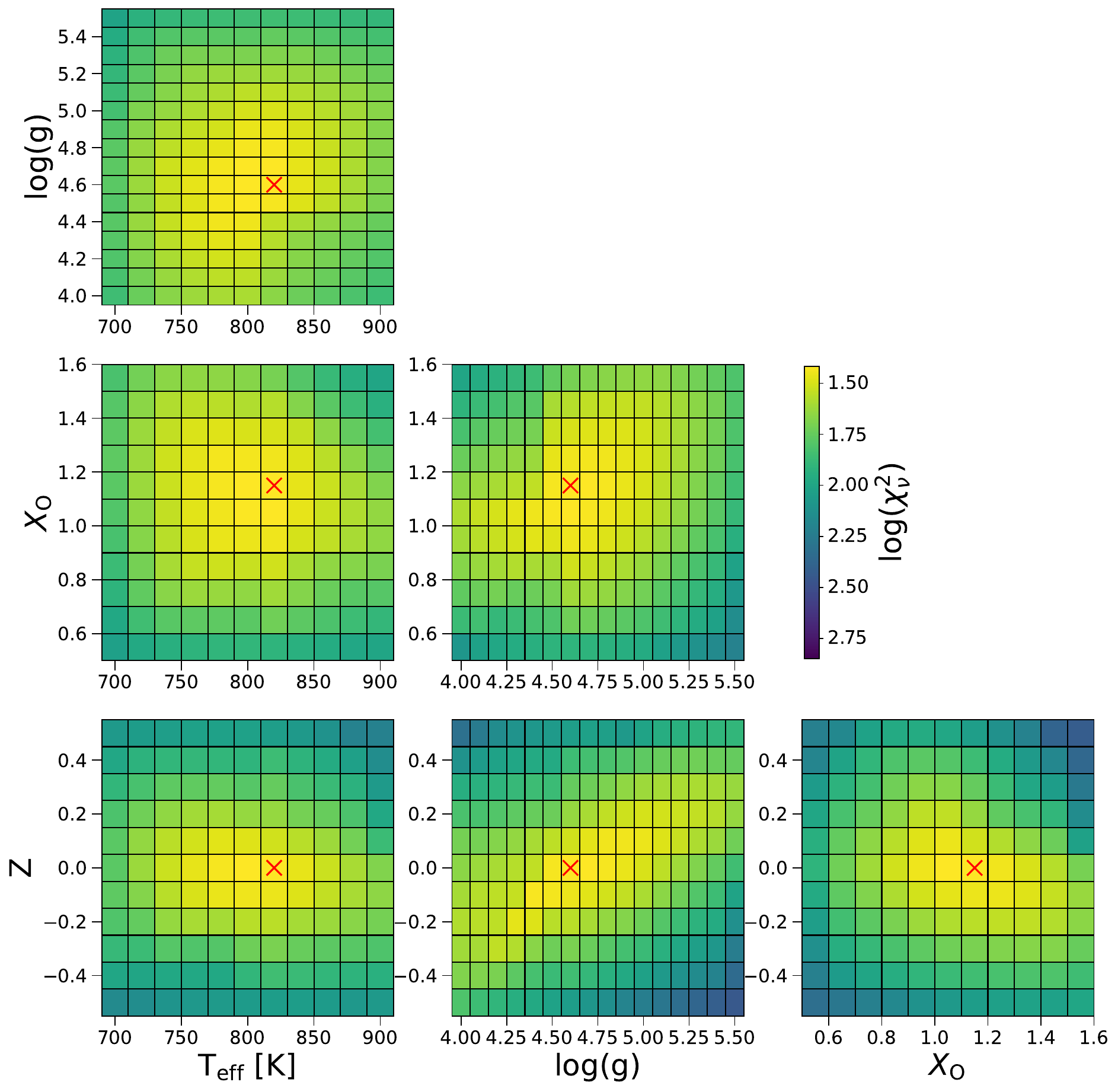}
\caption{GJ 570D $\chi_\nu^2$ surfaces for each parameter in the model grid. There are $\nu=3927$ degrees of freedom. The location of the best-fit value ($\chi^2_\nu=26.1$) is indicated by a red cross in each panel. \label{fig:GJ_570D_corner_plot}}
\end{figure}

The best-fit model spectrum to the GNIRS spectrum of GJ 570D is shown in the top panel of Figure \ref{fig:ALL_model_fit_fullspec}. We find the best-fit parameters to be $T_\mathrm{eff}=820\,$K, $\log(g)=4.6$, $Z=0.0$ and $\mathrm{C/O}=0.48$, using models with varying oxygen abundance. The best-fit parameters are well contained within the model grid parameter space, as shown by the $\chi^2$ surfaces in Figure \ref{fig:GJ_570D_corner_plot}. The minimum $\chi^2$ values are on the order of $10^5$ with 3927 degrees of freedom, and almost all the probability contained within the best-fit model. We also compare the best-fit models when fitting grids in which the C/O ratio is changed by altering either the carbon or oxygen elemental abundances, as shown in Figure \ref{fig:varO_varC_comparison}. We find negligible differences in the best-fitting model parameters, except for gravity and C/O ratio where there are small differences which do not impact the spectra meaningfully.  Therefore, in the grid-fitting analyses we use only models which alter the oxygen abundance of the model atmosphere. 

 We also fit the \texttt{ATMO} models to the $Y$, $J$, $H$ and $K$ bands of GJ 570D individually, as shown in Figure \ref{fig:GJ_570D_model_fit_zoomspec}. The individual-band fits reproduce the flux and shape of each band signficantly better than the full-spectrum model fit. However, the best-fit parameters are inconsistent and disagree with the results from the full-spectrum. The best-fit parameters of the full-spectrum and individual band fits are summarised in Table \ref{tab:GJ_570D_fit_table}.  

The age and metallicity of the GJ 570 system are constrained through observations of the host star GJ 570A (see Table 2 of \citealp{Zhang_2021_a}, and references therein). High resolution spectroscopy of the host star  indicates a metallicity of $Z=-0.05-0.05$. The host star C/O ratio has been found to be 0.65-0.97 from a stellar abundance analysis \citep{Line_2015}. Our best-fit metallicity is consistent with that of the host star. However our best-fit C/O ratio of 0.48 is lower compared to that of the host star ($\mathrm{C/O}=0.65-0.97$). The age of GJ 570A is in the range $1.4-5.2\,\mathrm{Gyr}$, determined from stellar isochrone fitting, gyrochronology and stellar activity.  Comparing our best-fit $T_\mathrm{eff}$, $\log(g)$ and $R$ to isochrones from the \texttt{ATMO} 2020 evolutionary models in Figure \ref{fig:evol_comparison}, we can see a disagreement with the values expected for an object with an age range of $1.4-5.2\,\mathrm{Gyr}$. Both the gravity and radius are underestimated compared to the evolutionary tracks.

%Our best-fit $T_\mathrm{eff}$ and $\log(g)$ imply a mass in the range $0.022-0.049\,M_\odot$ and an age in the range $0.09\to10.0\,$Gyr, using the \texttt{ATMO} 2020 evolutionary tracks \citep{Phillips_2020}.

Most previous grid fitting studies of GJ 570D used an IRTF/SpeX prism spectrum, providing $1.0-2.5\,\mu$m wavelength coverage and $R\approx150-400$. Fitting analyses have been done using forward modeling \citep{Oreshenko_2020, Zhang_2021_a} and retrieval frameworks \citep{Line_2015, Line_2017, Kitzmann_2020, Zalesky_2022, Whiteford_2023}, and the results from these studies compared to this work are shown in Table \ref{tab:GJ_570D_params}. We find a comparable effective temperature and higher gravity compared to \citet{Zhang_2021_a}, and consistent parameters compared to \citet{Oreshenko_2020}. Notably we find an approximate solar metallicity, whereas the forward modelling work of \citet{Zhang_2021_a} and the retrieval studies of \citet{Line_2015, Line_2017, Kitzmann_2020} all find sub-solar metallicities (see Table \ref{tab:GJ_570D_params}). Furthermore, our best-fit C/O ratio is approximately solar, whereas the retrieval studies find a super-solar C/O ratio (see Table \ref{tab:GJ_570D_params}). 

These differences could be due to differences in the modelling framework, or due to the higher resolution and signal-to-noise of the GNIRS spectrum. To test this we also fit the \texttt{ATMO} models to the IRTF/SpeX spectrum of GJ 570D \citep{Burgasser_2004}. We use the same methods as described in Section \ref{sec:methods} to degrade the model spectra. However to implement the wavelength-dependent resolution of the SpeX instrument ($R\sim150-400$, $0.8-2.5\,\mu$m, slit width $0.5"$) we vary the width of the Gaussian kernel used to smooth the spectra as described in \citet{Zhang_2021_a}. The best-fitting model is shown in Figure \ref{fig:ALLSpeX_model_fit}, and we can see that we obtain the same gravity and C/O ratio as the best-fitting model for the higher resolution GNIRS data. We obtain a cooler effective temperature of $T_\mathrm{eff}=800\,$K and a sub-solar metallicity of $Z=-0.1$. The higher resolution GNIRS data appears to better constrain the metallicity of this object, given that the metallicity is constrained to appoximately solar from observations of the host star. We also see that the same discrepancies between model and data exist in the best-fit to the SpeX observations, namely an underprediction of the peak flux in the $H$ band, underprediction of the flux in the red side of the $Y$ band, and an overprediction of flux on the blue side of the $J$ band.

\subsection{HD 3651B}
\label{sec:results_3651B}

\begin{figure}[ht!]
\plotone{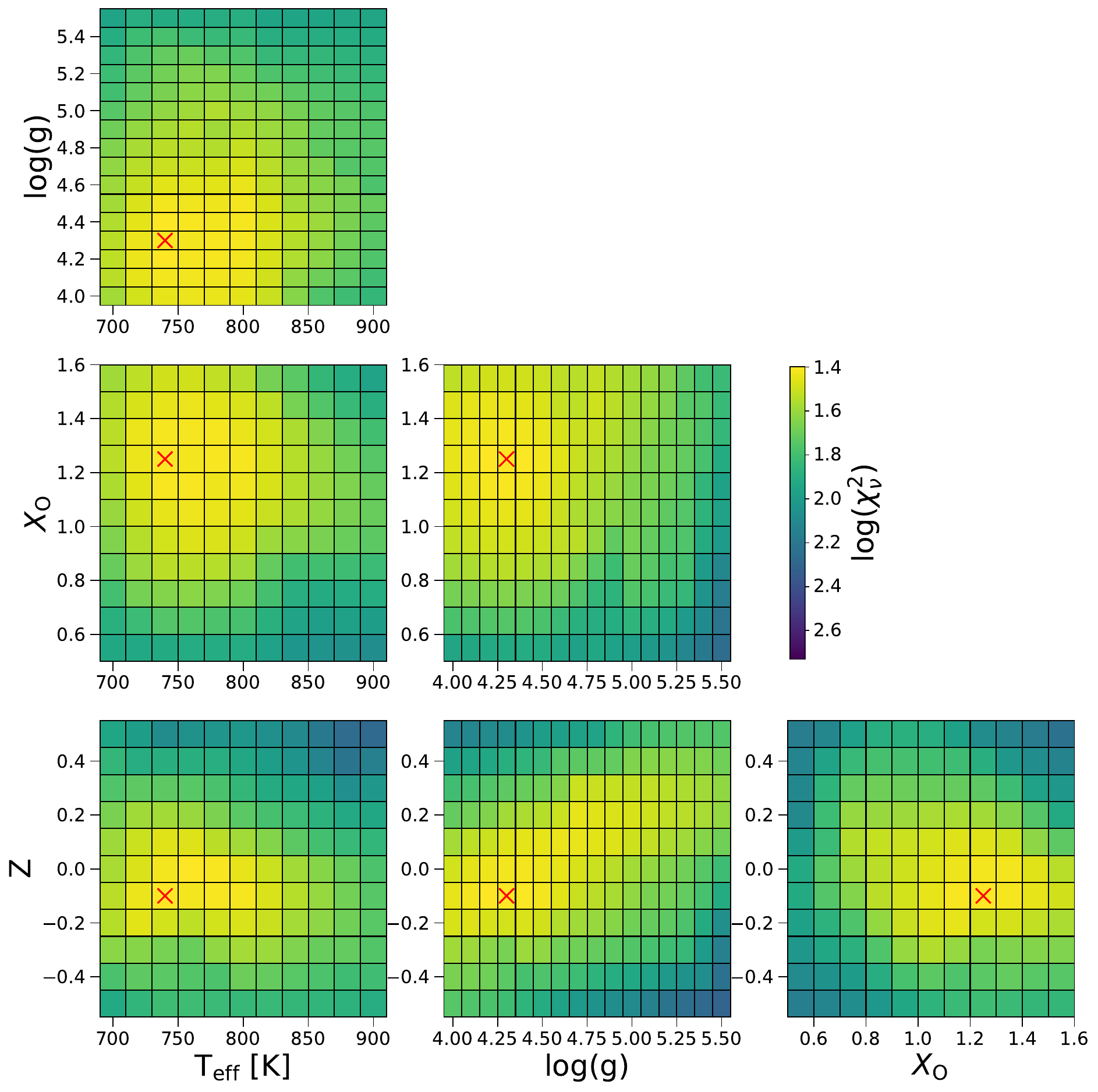}
\caption{HD 3651B $\chi_\nu^2$ surfaces for each parameter in the model grid. There are $\nu=3931$ degrees of freedom. The location of the best-fit value ($\chi^2_\nu=25.0$) is indicated by a red cross in each panel. \label{fig:HD3651B_corner_plot}}
\end{figure}

The best-fit model spectrum to the GNIRS spectrum of HD 3651B is shown in the middle panel of Figure \ref{fig:ALL_model_fit_fullspec}. We find the best-fit parameters to be $T_\mathrm{eff}=740\,$K, $\log(g)=4.3$, $Z=-0.1\,$ dex and $\mathrm{C/O}=0.44$, using models with varying oxygen abundance. The best-fit parameters are well contained within the model grid parameter space, as shown by the $\chi^2$ surfaces in Figure \ref{fig:HD3651B_corner_plot}. The minimum $\chi^2$ values are on the order of $10^5$ with 3923 degrees of freedom. 

In addition to fitting the full wavelength range, we also fit the Y, J, H and K bands individually as shown in Figure \ref{fig:HD3651B_model_fit_zoomspec}. Similar to GJ 570D, fitting the individual bands alleviates the model-data discrepancies seen in the full wavelength range best-fit model, and better reproduces the flux and shape of each band. However, the best-fit parameters are again inconsistent between the individual bands and disagree with the results from the full-spectrum. The best-fit parameters and wavelength ranges of the individual band fits are summarized in Table \ref{tab:HD_3651B_fit_table}. 

The age and metallicity of HD 3651B are constrained through observations of the K0V host star HD 3651A \citep{Liu_2007, Zhang_2021_a}. High resolution spectroscopy of HD 3651A indicates a metallicity of $Z=0.1-0.2\,$dex, and the host star C/O ratio has been found to be $0.51-0.73$ from a stellar abundance analysis \citep{Line_2015}. Both our best-fit metallicity and C/O ratio are subsolar and below that of the host star, though consistent within our adopted uncertainties of half the model grid spacing. The age of the HD 3651 system is in the range $4.5-8.3\,$Gyr, determined from stellar isochrone fitting, gyrochronology and stellar activity \citep{Zhang_2021_a}. We compare our best-fit $T_\mathrm{eff}$, $\log(g)$ and $R$ to that expected for HD 3651B from \texttt{ATMO} 2020 model isochrones. Similar to GJ 570D, the gravity is underpredicted compared to that expected for a $4.5-8.3\,$Gyr object. However unlike GJ 570D, the radius is overestimated. 

\begin{figure}[ht!]
\plotone{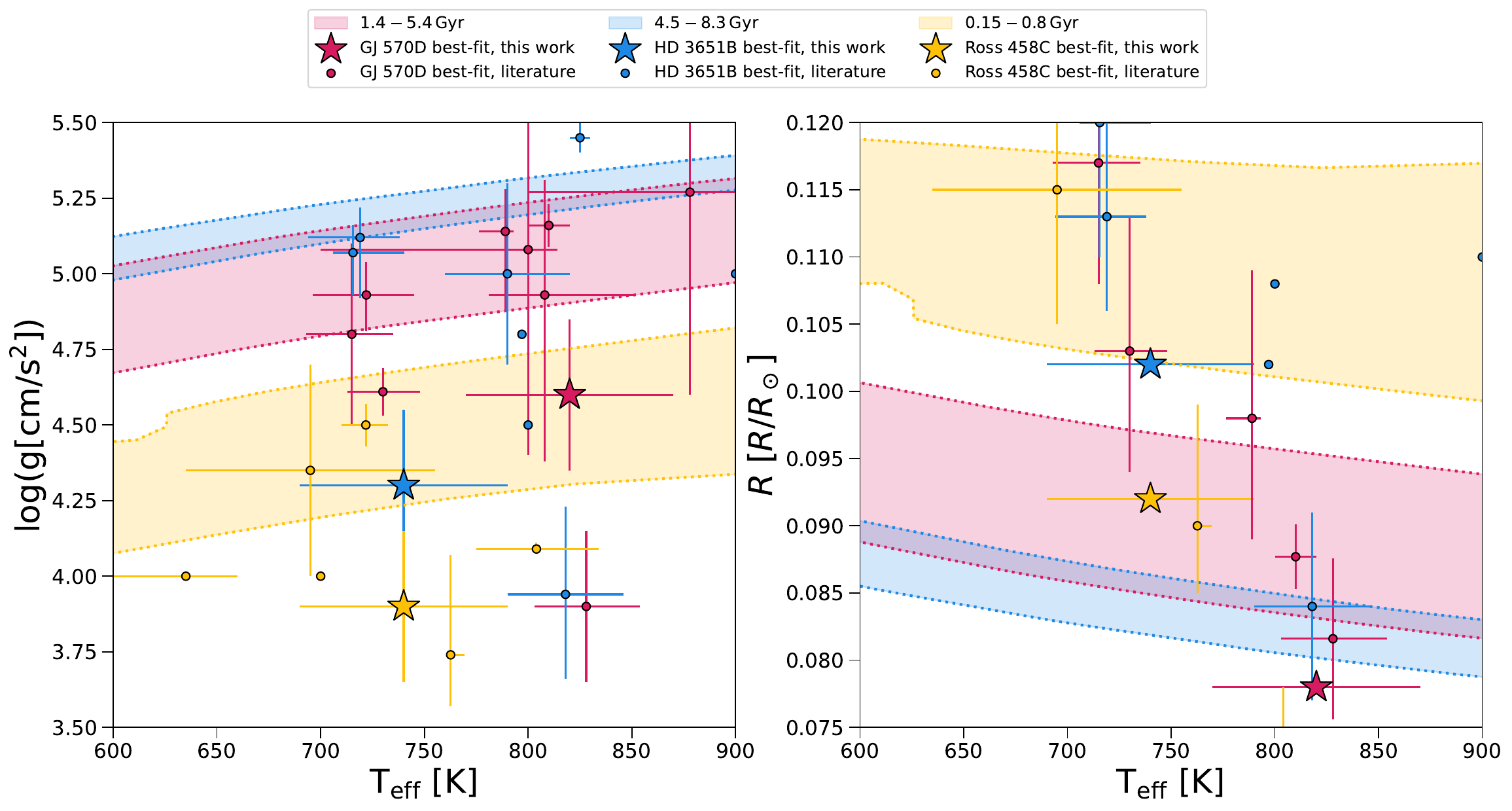}
\caption{ Comparison of our best-fit gravity, effective temperature and radius to the \texttt{ATMO} 2020 evolutionary tracks \citep{Phillips_2020}. The best-fit values to the GNIRS spectrum of each object are shown as the colored stars with error bars. The best-fit values from spectroscopic analyses in the literature listed in Tables \ref{tab:GJ_570D_params}, \ref{tab:HD3651B_params} and \ref{tab:ROSS_458C_params} are shown as smaller color circles with error bars. Note that some literature sources do not provide best-fit radii.  The corresponding colored shaded regions are the range of estimated ages of each benchmark system, with the dotted lines as model isochrones. \label{fig:evol_comparison}}
\end{figure}

HD 3651B has been the subject of numerous previous grid fitting studies using its IRTF/SpeX prism spectrum. \citet{Burgasser_2007} used a semi-empirical method comparing measured spectral indices to spectral models that were calibrated on GJ 570D to determine the physical properties. \citet{Leggett_2007} also obtained a GNIRS spectrum of HD 3651B for comparisons with atmosphere models, and \citet{Liu_2007} fit the SpeX spectrum of HD 3651B to several model grids. Most recently, fitting analyses on the SpeX spectrum have been performed using forward modelling \citep{Zhang_2021_a} and retrieval frameworks \citep{Line_2015, Line_2017, Zalesky_2022}. The results from these studies compared to this work are shown in Table \ref{tab:HD3651B_params}. We find a lower $T_\mathrm{eff}$ compared to the \texttt{Sonora} model fits of \citet{Zhang_2021_a}, but a similar surface gravity. The retrieval analyses obtain a larger surface gravity (see Table \ref{tab:ROSS_458C_params}) more in line with that expected from evolutionary models for this object (see Figure \ref{fig:evol_comparison}), and effective temperatures comparable to that found in this work. 

We also fit the \texttt{ATMO} models to the IRTF/SpeX spectrum of HD 3651B \citep{Burgasser_2007}, and the best-fitting model is shown in Figure \ref{fig:ALLSpeX_model_fit}. We find a larger $T_\mathrm{eff}$ of $800\,$K compared to the GNIRS model fit ($T_\mathrm{eff}=740\,K$), more in line with the $T_\mathrm{eff}$ obtained in \citet{Zhang_2021_a} and older fitting analyses \citep{Burgasser_2007, Leggett_2007, Liu_2011}. The gravity, however, is comparable to the GNIRS model fit, and remains low compared to evolutionary models. The metallicity increases by $0.1\,$dex, but is still low compared to the metallicity of the host star. The C/O ratio does not differ significantly. Interestingly, the radius of the SpeX model fit is lower compared to the GNIRS model fit, and is consistent with that predicted by the evolutionary models for this object (see Figure \ref{fig:evol_comparison}). This occurs due to the higher effective temperature of the SpeX model fit allowing a lower radius to be used when fitting the data.

\subsection{Ross 458C}
\label{sec:results_458C}

\begin{figure}[ht!]
\plotone{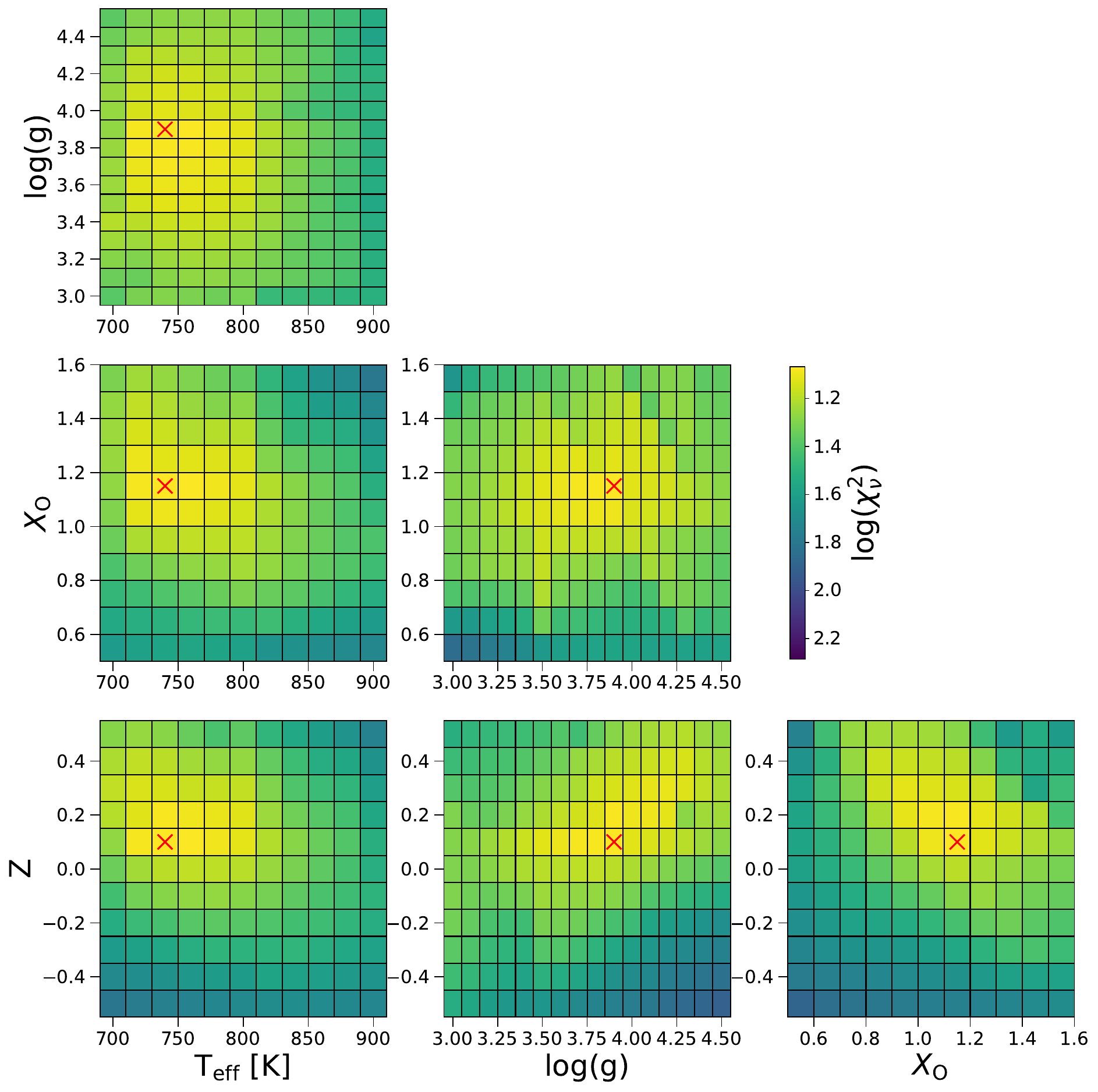}
\caption{Ross 458C $\chi_\nu^2$ surfaces for each parameter in the model grid. There are $\nu=3907$ degrees of freedom. The location of the best-fit value ($\chi^2_\nu=11.7$) is indicated by a red cross in each panel. \label{fig:ROSS_458C_corner_plot}}
\end{figure}

The best-fit model spectrum to the GNIRS spectrum of Ross 458C is shown in Figure \ref{fig:ALL_model_fit_fullspec}. We find the best-fit parameters to be $T_\mathrm{eff}=740\,$K, $\log(g)=3.9$, $Z=0.1\,$dex and $\mathrm{C/O}=0.48$, using models with varying oxygen abundance. The best-fit parameters are well contained within the model grid parameter space. This is shown in the $\chi^2$ surfaces in Figure \ref{fig:ROSS_458C_corner_plot}. The minimum $\chi^2$ values are on the order of $10^5$ with 3907 degrees of freedom.  

 Similarly to GJ 570D and HD 3651B, we also fit the \texttt{ATMO} models to the $Y$, $J$, $H$ and $K$ bands of Ross 458C individually, as shown in Figure \ref{fig:ROSS_458C_model_fit_zoomspec}. As with the previous two objects, the individual-band fits reproduce the flux and shape of each band significantly better than the full spectrum model fit. The best-fit parameters are again inconsistent and disagree with the results from the full-spectrum. The best-fit parameters of the full spectrum and individual-band fits are summarised in Table \ref{tab:Ross_458C_fit_table}.

%The regions of the near-infrared spectrum that are not fitted satisfactorily are shown in more detail in Figure \ref{fig:ROSS_458C_model_fit_zoomspec}. These are the same wavelength regions and issues as seen in the best-fit model of GJ 570D (Figures \ref{fig:GJ_570D_model_fit_fullspec} and \ref{fig:GJ_570D_model_fit_zoomspec}), namely the model underpredicts the flux in the red side of the $Y$ band ($\approx1.07-1.10\,\mu$m), overpredicts the flux on the blue side of the $J$ band ($\approx1.22-1.26\,\mu$m), and underpredicts the peak flux of the $H$ band ($\approx1.54-1.60\,\mu$m). Unlike the best-fit model of GJ 570D, the best-fit model of Ross 458C also slightly underpredicts the peak flux in the $K$ band ($\approx2.05-2.12\,\mu$m).  

The age and metallicity of the Ross 458 system are constrained through observations of the host binary Ross 458AB. Photometry and medium-resolution ($R\approx2000$) spectroscopy of Ross 458A indicate a super-solar metallicity in the range $Z=0.17-0.33\,$dex \citep{Burgasser_2010, Gaidos_Mann_2014}.  Our best-fit metallicity of Ross 458C is super-solar, but slightly lower than that observed for the host star at $Z=+0.1\,$dex.  The age of Ross 458A is in the range $0.15-0.8\,$Gyr, which has been determined from gyrochronology, stellar activity, and spectroscopic youth indicators \citep{Burgasser_2010, Zhang_2021_a}.  We compare our best-fit $T_\mathrm{eff}$, $\log(g)$ and $R$ to that expected for Ross 458C from \texttt{ATMO} 2020 model isochrones. Similarly to both GJ 570D and HD 3651B, the gravity of Ross 458C is underpredicted compared to that expected for a $0.15-0.8\,$Gyr object. Additionally the radius of Ross 458C is also lower compared to the evolutionary models. 

Ross 458C has been studied extensively in the literature using near-infrared spectra from multiple different spectrographs. \citet{Burgasser_2010} and \citet{Morley_2012} used a low-resolution ($R\sim250-350$) Magellan/FIRE spectrum of Ross 458C and cloudy atmosphere models to derive its physical parameters. \citet{Burningham_2011} obtained a Subaru/IRCS spectrum and, combined with photometry and age-estimates of the system, estimated the bolometric luminosity and physical parameters of Ross 458C. Most recently, fitting analyses of the IRTF/SpeX prism spectrum of Ross 458C have been performed using forward modelling \citep{Zhang_2021_a} and retrieval frameworks \citep{Zalesky_2022}. \citet{Gaarn_2023} also performed a retrieval analysis on the Magellan/FIRE spectrum of Ross 458C. The results from these studies compared to this work are shown in Table \ref{tab:Ross_458C_fit_table}. We find a comparable effective temperature and surface gravity to the retrieval studies, and note that the higher gravity retrieved by \citet{Gaarn_2023} is more consistent with that expected from evolutionary models. We obtain a comparable gravity ($\log(g)=3.9$) but lower effective temperature ($T_\mathrm{eff}=740\,$K) compared to the \texttt{Sonora} forward modelling work of \citet{Zhang_2021_a}. Furthermore, we get a lower metallicity ($Z=0.1\,$dex) than the \texttt{Sonora} models obtained on the SpeX spectrum. 

We also fit our \texttt{ATMO} models to the IRTF/SpeX spectrum of Ross 458C \citep{Zhang_2021_a}. The best-fitting model is shown in Figure \ref{fig:ALLSpeX_model_fit}, and we obtain notably different best-fitting parameters compared to our higher-resolution GNIRS data. This is due to a large discrepancy between the observed GNIRS and SpeX spectra of Ross 458C, with the $H$ and $K$ bands in the SpeX spectrum having a larger flux in the normalised spectra. The near-infrared colors of the SpeX spectrum are therefore redder than the GNIRS spectrum for Ross 458C, with $J-H=-0.18$ and $J-K=0.31$ for the SpeX spectrum, and $J-H=-0.38$ and $J-K=0.19$ for the GNIRS spectrum. Near-infrared spectra of Ross 458C have been obtained using IRTF/SpeX in July 2015 \citep{Zhang_2021_a}, Magellan/FIRE in April 2010 \citep{Burgasser_2010} and Subaru IRCS in May and December 2009 \citep{Burningham_2011}. Comparing these spectra to our Gemini/GNIRS spectrum in Figure \ref{fig:GNIRS-SpeX_Ross458C_comp}, we can see differences in the $J-H$ and $J-K$ colors. While we see variability in these colors, we do not see any variability in the shape of the individual bands.  It should be noted that these spectral differences could be caused by imperfect order stitching in the GNIRS and IRCS data. However the SpeX and FIRE spectra also show slight differences in the $J-H$ and $J-K$ colors, and are not impacted by order stitching since these spectra were obtained in prism mode, which produces continuous wavelength coverage. 

Ross 458C has been previously observed to be variable by \citet{Manjavacas_2019} who obtained time-resolved spectroscopy using the Hubble Space Telescope Wide Field Camera 3. Over their $\sim10\,$h of observing time they found Ross 458C to have peak-to-peak rotational modulations of $2.63\pm0.02\%$ over the entire $1.1-1.64\,\mu$m wavelength range, with an estimated $6.75\pm1.58\,$h rotation period. This has been attributed to heterogeneous sulfide clouds in the atmosphere. The peak-to-peak variability amplitude is smaller than seen in our comparisons of the spectra of Ross 458C in Figure \ref{fig:GNIRS-SpeX_Ross458C_comp}. It should also be noted that \citet{Metchev_2015} did not detect this object to be variable in the \textit{Spitzer} IRAC channels 1 ($3.6\mu$m) and 2 ($4.5\,\mu$m), observing variability amplitudes of $<1.37\%$ and $<0.72\%$ over time baselines of 14 and $7\,$h respectively. 

\section{Discussion}
\label{sec:discussion}

There are three regions of the near-infrared spectrum that are not fitted satisfactorily, and these regions are consistent for the best fits of all three objects. The best-fit models underpredict the flux in the red side ($\approx1.07-1.10\,\mu$m) of the $Y$ band, overpredict and underpredict the flux on the blue ($\approx1.22-1.26\,\mu$m) and red ($\approx1.27-1.31\,\mu$m) side of the $J$ band, respectively, and underpredict the peak flux ($\approx1.54-1.60\,\mu$m) of the $H$ band. This is seen in the spectral-fitting residuals shown in Figure \ref{fig:benchmark_residuals}. These discrepancies are alleviated when fitting to the $Y$, $J$, $H$ and $K$ bands individually (Figures \ref{fig:GJ_570D_model_fit_zoomspec}, \ref{fig:HD3651B_model_fit_zoomspec} and \ref{fig:ROSS_458C_model_fit_zoomspec}).  These individual-band fits reproduce the flux and shape of each band significantly better than the full-spectrum fit, indicating that deficiencies lie in the temperature profile and/or the corresponding chemical abundances. For example, the issues could lie in our assumption of radiative-convective equilibrium, or in using a constant $K_{zz}$ in the convective and radiative regions of the atmosphere \citep[e.g.][]{Mukherjee_2022}. Furthermore, potential uncertainties in the order stitching of the GNIRS data by \texttt{PypeIt} could introduce uncertainties to the baseline of the spectral shape. We now discuss each of the $Y$, $J$, $H$ and $K$ bands in turn, beginning with the $H$ band. 

\subsection{H band}

The peak flux in the $H$ band is underestimated in the best-fit models of GJ 570D, HD 3651B and Ross 458C. The individual spectral band fits better reproduce the flux in the $H$ band, and the reduced $\chi^2$ values are lower in the individual spectral band fits than the full-spectrum fits (see Tables \ref{tab:GJ_570D_fit_table}, \ref{tab:HD_3651B_fit_table} and \ref{tab:Ross_458C_fit_table}). By comparing the best-fit parameters of the individual $H$ band fits to those of the full-spectrum fits, we can see a clear trend. The best-fit models move to higher gravity and higher metallicity in order to better reproduce the data. To examine why this is the case, we compare the properties at the photosphere (pressure, temperature, abundances) in the $H$ band, for models with $\log(g)=4.5$, $Z=0.0$ and models with $\log(g)=5.5$, $Z=0.5$. 

\begin{figure}[ht!]
\plotone{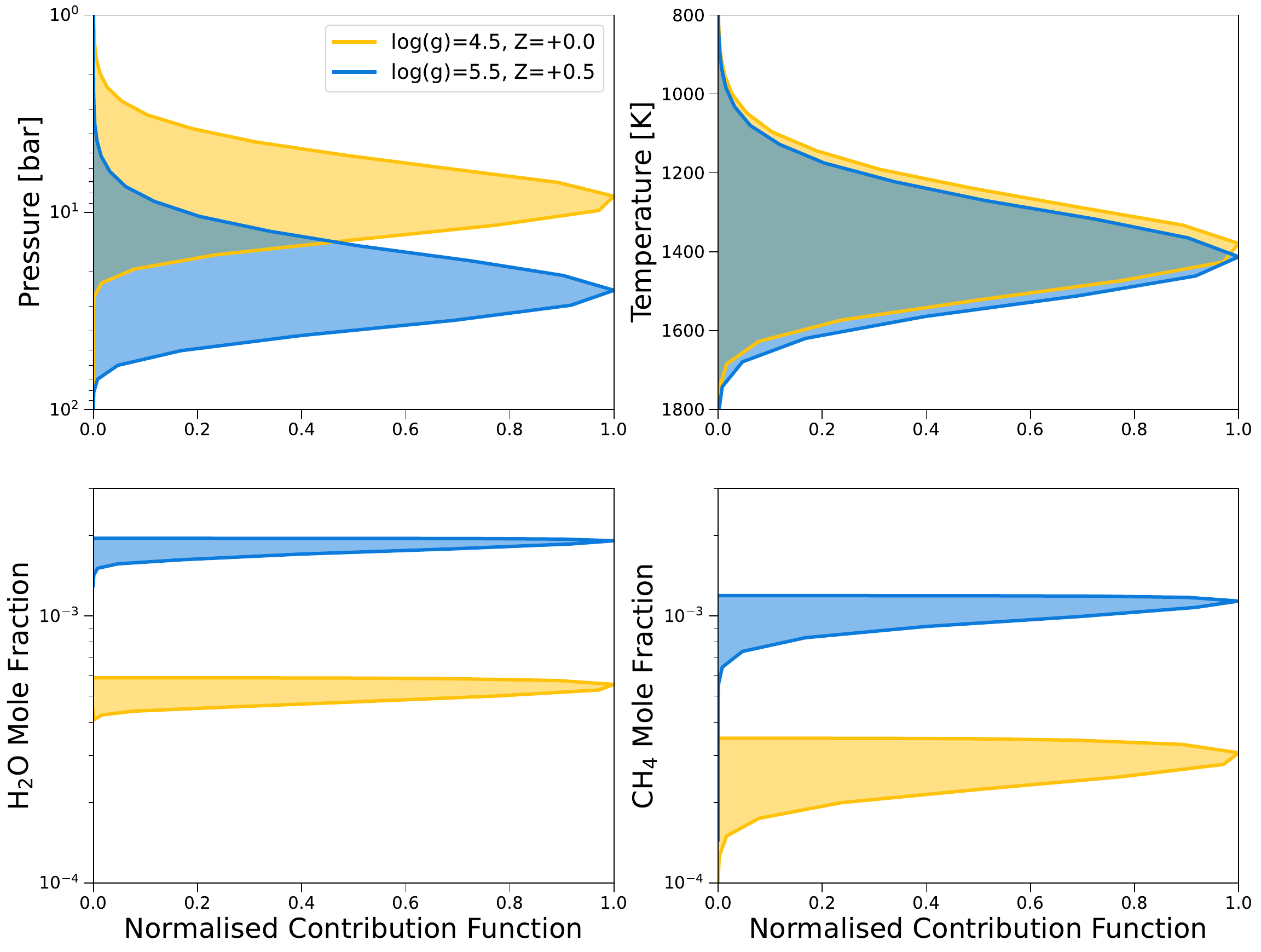}
\caption{Atmospheric pressure, temperature, $\mathrm{H_2O}$ and $\mathrm{CH_4}$ mole fractions (clockwise from top left), for the $1.58\,\mu$m contribution function for a low-gravity solar metallicity atmosphere (yellow) and a high-gravity super-solar metallicity atmosphere (blue) as indicated in the legend. Increasing gravity and metallicity shifts the H-band photosphere to higher pressure and increases the photospheric abundances of important opacity sources $\mathrm{H_2O}$ and $\mathrm{CH_4}$. \label{fig:contribution_function}}
\end{figure}

We first begin by examining the contribution function at a given wavelength. The pressure and temperature along with the $\mathrm{H_2O}$ and $\mathrm{CH_4}$ mole fractions for the $1.58\,\mu$m contribution function are shown in Figure \ref{fig:contribution_function}. It should be noted that these are the properties for the contribution function at a single wavelength, and there will be further variation of these properties if a larger wavelength range is selected (for example, the entirety of the $H$ band). In Figure \ref{fig:contribution_function} we can see that in high-gravity, high-metallicity atmospheres, the contribution function moves to higher pressures and stays at approximately the same temperature. This is the result of two competing effects. Increasing the gravity of a model atmosphere shifts the contribution to higher pressures due to a decrease in the pressure scale height and the optical depth at a given pressure level. At a given pressure level, the temperature decreases with increasing gravity. Conversely, increasing the metallicity of a model atmosphere increases the abundances of metal-bearing species such as $\mathrm{H_2O}$ and $\mathrm{CH_4}$, increasing the opacity of the atmosphere and shifting the contribution to lower pressures and higher temperatures. In the case shown in Figure \ref{fig:contribution_function}, the contribution function shifts to higher pressures and remains at approximately the same temperature when increasing gravity and metallicity. Figure \ref{fig:contribution_function} also shows that the abundances of important opacity sources $\mathrm{H_2O}$ and $\mathrm{CH_4}$ increase at the location of the contribution function, primarily due to the increase in metallicity. 

Figure \ref{fig:Hband_model_spec_analysis} shows how the H-band model spectra change when increasing the gravity and metallicity of the atmosphere. At $\log(g)=4.5$ and $Z=0.0$ (i.e. parameters representative of the best-fit model to the full spectra of the objects), increasing either the gravity or metallicity does little to increase the peak H-band flux and reproduce the observations. However, it can be seen at high gravities and metallicities, increasing these parameters increases the peak flux in the $H$ band. This shows that the effects gravity/metallicity have on a spectrum is dependent on the metallicity/gravity of the atmosphere. To understand why this is the case, Figure \ref{fig:Hband_model_kabs_analysis} shows the opacities shaping the $H$ band spectrum at the location in the model atmosphere of the peak of the contribution function. The blue side of the $H$ band is shaped by $\mathrm{H_2O}$ opacity, and the red side by $\mathrm{CH_4}$ opacity. The peak flux is controlled by these two opacities sources, along with $\mathrm{H_2-H_2}$ collisionally induced absorption, the opacity of which depends on gravity and metallicity. Increasing gravity increases $\mathrm{H_2-H_2}$ CIA due to the higher atmospheric pressures. Increasing metallicity however, decreases the $\mathrm{H_2-H_2}$ CIA due to more hydrogen being sequestered into molecules such as $\mathrm{H_2O}$, $\mathrm{CH_4}$ and $\mathrm{NH_3}$. The abundances of these molecules increase due to the increased availability of metals, increasing the opacity of these species and decreasing $\mathrm{H_2-H_2}$ CIA. The interplay/competition of the effects of gravity and metallicity on $\mathrm{H_2-H_2}$ CIA is what drives the individual band fits to converge on high gravity, high metallicity models. These parameters allow the model to control the peak $H$ band flux through the level of $\mathrm{H_2-H_2}$ CIA. 

While these model fits reveal that the peak $H$-band flux can be affected by the level of $\mathrm{H_2-H_2}$ CIA, in cooler atmospheres the abundance of $\mathrm{NH_3}$ has also been shown to affect the flux emitted through the $H$ band. The abundance of $\mathrm{NH_3}$ increases in cooler atmospheres, leading to increased opacity in the $H$ band. However, non-equilibrium chemistry models are required to quench the abundance of $\mathrm{NH_3}$ through vertical mixing, increasing the flux in the $H$-band compared to chemical equilibrium models that is required to reproduce observations of late-T and Y dwarfs \citep{Tremblin_2015}. Since we include non-equilibrium chemistry due to vertical mixing in the \texttt{ATMO} models, the continued under-prediction of the flux in the $H$ band could indicate that the abundance of $\mathrm{NH_3}$ needs to be further reduced, implying inadequacies in the chemistry model. To explore this, we removed $\mathrm{NH_3}$ opacity from the calculation of the emission spectra for models close to the best-fitting parameters of GJ 570D and Ross 458C. We note that this treatment is not self-consistent since we do not recompute the T-P profile. The result is shown in Figure \ref{fig:noNH3_comparison}, in which we can see a H-band brightening in the models without $\mathrm{NH_3}$ opacity which becomes larger when decreasing the effective temperature. However, while removing the $\mathrm{NH_3}$ opacity does brighten the $H$ band, it does not fully account for the under-prediction of flux seen in our best-fits of GJ 570D and Ross 458C.

\subsection{Y band}

 The best-fit models for all objects underestimate the flux in the red-side of the $Y$-band ($\approx1.07-1.10\,\mu$m). The best-fit model of Ross 458C also overestimates the flux in the blue side of the $Y$-band. These discrepancies can be seen in the scaled residuals shown in Figure \ref{fig:benchmark_residuals}.  The individual band fits rectify these issues for all objects, and we can see a similar trend as $H$ band, in that the best individual band fit models move to higher gravity and metallicity in order to better reproduce the data (see Tables \ref{tab:GJ_570D_fit_table}, \ref{tab:HD_3651B_fit_table} and \ref{tab:Ross_458C_fit_table}).

%The pressure and temperature, along with the $\mathrm{H_2O}$ and $\mathrm{CH_4}$ mole fractions at the $1.09\,\mu$m contribution function are shown in Figure \ref{fig:Yband_contribution_function}.  Similarly to the H band, the Y band contribution function moves to higher pressures, and approximately $\sim80\,$K warmer temperatures when increasing the gravity and metallicity. The abundances of $\mathrm{H_2O}$ and K also increase at the location of contribution function when increasing these parameters, primarily due to the increase in metallicity. 

Figure \ref{fig:Yband_model_spec_analysis} shows how the Y-band model spectra change when increasing the gravity and metallicity of the atmosphere. At $\log(g)=4.5$ and $Z=0.0$, increasing the metallicity does not increase the peak Y-band flux, or alter the shape of the Y-band significantly. However, increasing the metallicity at high gravity ($\log(g)=5.5$), increases the peak flux of the $Y$ band. Increasing the gravity, regardless of the metallicity, leads to an increase in the flux passing through the $Y$ band, with a larger increase seen at solar metallicity. Figure \ref{fig:Yband_model_kabs_analysis} shows the opacities shaping the $Y$ band spectrum at the location of the peak of the contribution function. The peak flux of the $Y$ band is primarily controlled by the red wing of the potassium resonance doublet, with $\mathrm{H_2O}$ opacity defining the red side of the band. At lower gravities, FeH contributes to the $Y$ band opacity, however increasing gravity lowers its abundance due to the formation of Fe-bearing condensates leading to the rainout of this element. Increasing gravity also leads to an increase in the opacity of potassium and $\mathrm{H_2-H_2}$ CIA,   which becomes the second most important absorber in the $Y$-band.  This is due to the increase in pressure at the peak of the contribution function, suppressing the flux in the $Y$ band. Increasing metallicity, further increases the potassium opacity along with $\mathrm{H_2O}$ opacity, but decreases the $\mathrm{H_2-H_2}$ CIA due to more hydrogen being contained within metal-bearing molecules. The combination of increasing the potassium opacity by increasing the gravity, and subsequently decreasing the $\mathrm{H_2-H_2}$ CIA by increasing the metallicity, reshapes the $Y$ band such that the individual band fits converge on a higher-gravity, higher-metallicity model. 

\subsection{J band}

The full-spectrum fits of all three benchmark objects do not correctly reproduce the shape of the $J$-band. They suffer from the same deficiencies; an overprediction of flux on the blue ($\approx1.22-1.26\,\mu$m) side of the band and an underprediction of flux on the red ($\approx1.27-1.31\,\mu$m) side of the band. This is seen in the scaled spectral-fitting residuals in Figure \ref{fig:benchmark_residuals}. When performing individual spectral band fits, the shape of the $J$-band is better reproduced by the best-fit models, as shown in Figures \ref{fig:GJ_570D_model_fit_zoomspec}, \ref{fig:HD3651B_model_fit_zoomspec} and \ref{fig:ROSS_458C_model_fit_zoomspec}. In these individual band fits, the best-fit models have different parameters for all three objects (Tables \ref{tab:GJ_570D_fit_table}, \ref{tab:HD_3651B_fit_table} and \ref{tab:Ross_458C_fit_table}). The best-fit model for GJ 570D shifts to lower $T_\mathrm{eff}$, higher gravity, and a super-solar metallicity and C/O ratio. The best-fit parameters for HD 3651B shift to higher gravity (edge of the model grid), a more super-solar metallicity, and C/O ratio while remaining at the same $T_\mathrm{eff}$ as the full spectrum fit. Finally, the best-fit parameters for Ross 458C follow a different trend, shifting to a lower gravity, sub-solar metallicity and C/O ratio while remaining at the same effective temperature. We can therefore see that the individual $J$-band fits show a similar trend to the $H$ and $Y$ bands, in that the best-fit parameters shift to high gravity and metallicity (except for Ross 458C).

Figure \ref{fig:Jband_model_spec_analysis} shows how the J-band model spectra change when increasing gravity and metallicity in the atmosphere. At $\log(g)=4.5$ and $Z=0.0$, increasing the metallicity does not impact the $J$ band. However increasing the metallicity at high gravity ($\log(g)=5.5$) increases the peak $J$-band flux and alters the shape of the band. Increasing gravity at solar and super-solar metallicities does not significantly impact the $J$-band, apart from the depth of the K I resonance doublet at $\approx1.25\,\mu$m which is strongest at $\log(g)=4.5$, $Z=+0.5$. Figure \ref{fig:Jband_model_kabs_analysis} shows the opacities shaping the $J$-band spectrum at the location of the peak of the contribution function. The $J$ band is primarily shaped by $\mathrm{H_2O}$ absorption, with potassium resonance doublets ($\approx1.175\,\mu$m and $\approx1.250\,\mu$m) and $\mathrm{H_2-H_2}$ CIA contributing to the opacity. At high gravities, the strength of $\mathrm{H_2-H_2}$ CIA increases as the contribution function shifts to higher pressures, and the $K$ I resonance doublets decrease in strength as this element begins to rainout of the atmosphere due to condensation. Increasing metallicity strengthens the $K$ resonance doublets, and decreases $\mathrm{H_2-H_2}$ CIA. The individual $J$-band fits of GJ 570D and HD 3651B therefore shift to higher gravity, higher metallicity atmospheres to better reproduce the shape of the $J$ band through these opacity sources. We also note that decreasing the effective temperature (as in the case of the GJ 570D best-fit parameters) also alters the shape of the $J$-band due to an increase in strength of the water absorption bands, (Figure \ref{fig:Jband_models}).

\subsection{K band}

The shape and flux level in the K band is well reproduced by the best-fit models for all objects. The peak flux of the $K$-band is slightly underestimated in the best-fit model of HD 3651B, and the best-fit model of Ross 458C slightly underestimates the flux on the blue side of the $K$-band. The individual spectral band fits visually do a better job at reproducing the depth of the water absorption features in the blue side of the $K$ band for Ross 458C and the peak flux for HD 3651B.

\section{Summary and Conclusions} \label{sec:conclusions}

\begin{figure}[ht!]
\begin{center} 
\plotone{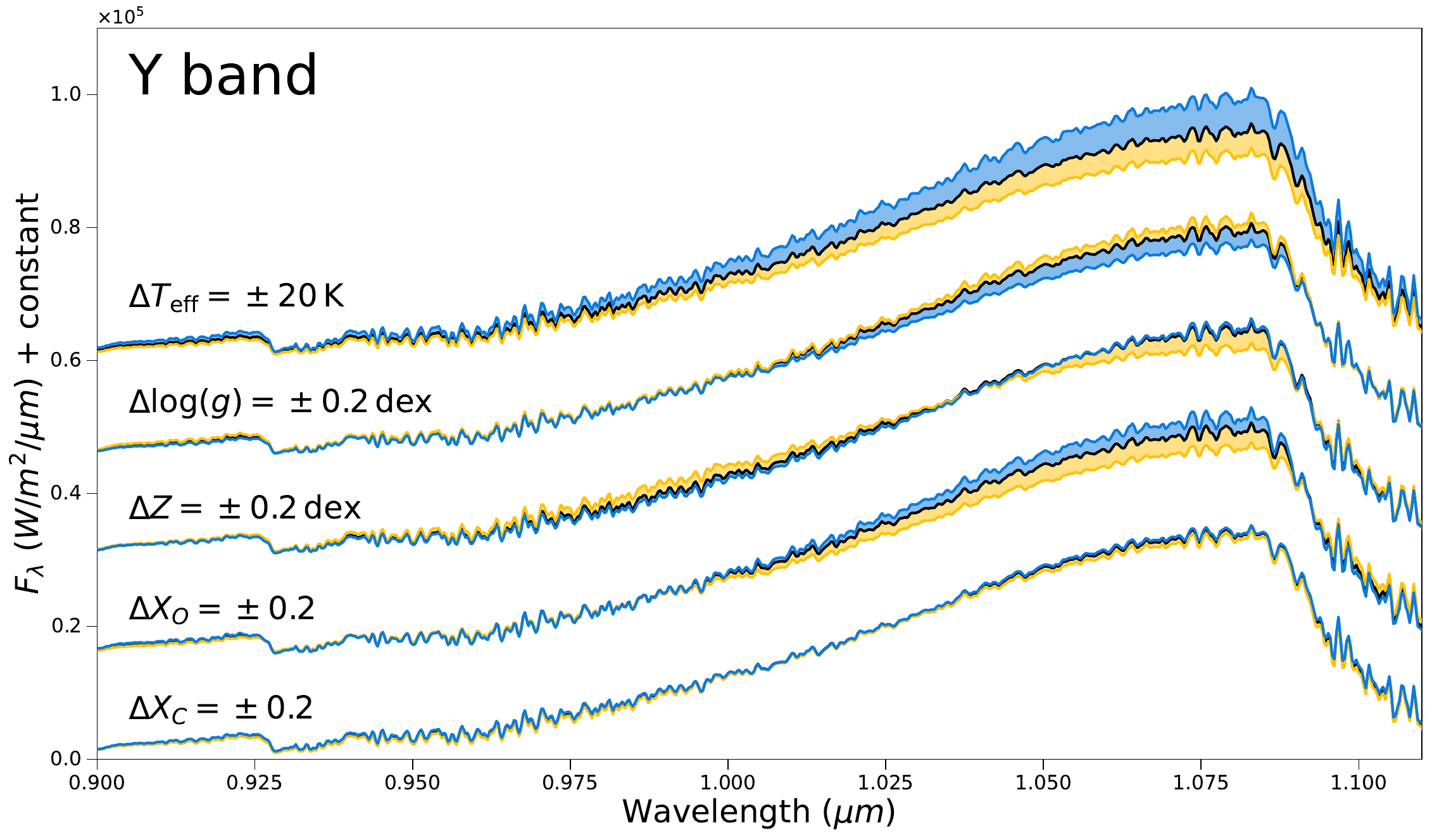}
\end{center}
\caption{Model \texttt{ATMO} $Y$-band emission spectra varying parameters around a $T_\mathrm{eff}=800\,K$, $\log(g)=5.0$, $Z=0.0\,$dex and $C/O=0.55$ (solar) model atmosphere. The oxygen and carbon abundances are changed through the scaling factors $X_O$ and $X_C$ respectively. Changes in the flux when increasing and decreasing the model grid parameters by the values indicated on the plot, are plotted as blue and yellow shaded regions respectively. The spectra have been degraded to the GNIRS resolution $R=1700$, and the models have been offset for clarity. \label{fig:Yband_models}}
\end{figure}

\begin{figure}[ht!]
\begin{center} 
\plotone{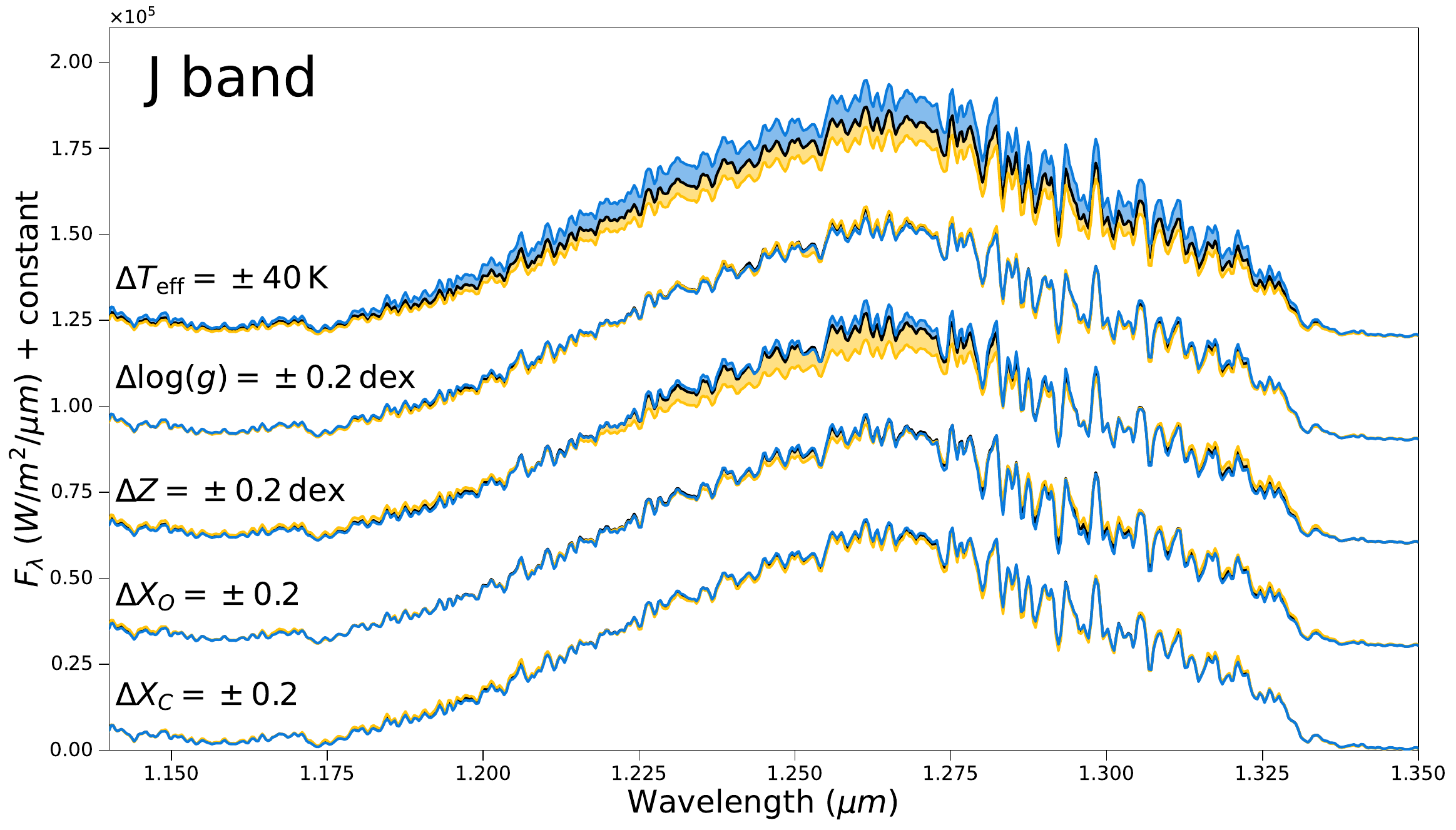}
\end{center}
\caption{Same as Figure \ref{fig:Yband_models} but for the $J$ band. \label{fig:Jband_models}}
\end{figure}

\begin{figure}[ht!]
\begin{center} 
\plotone{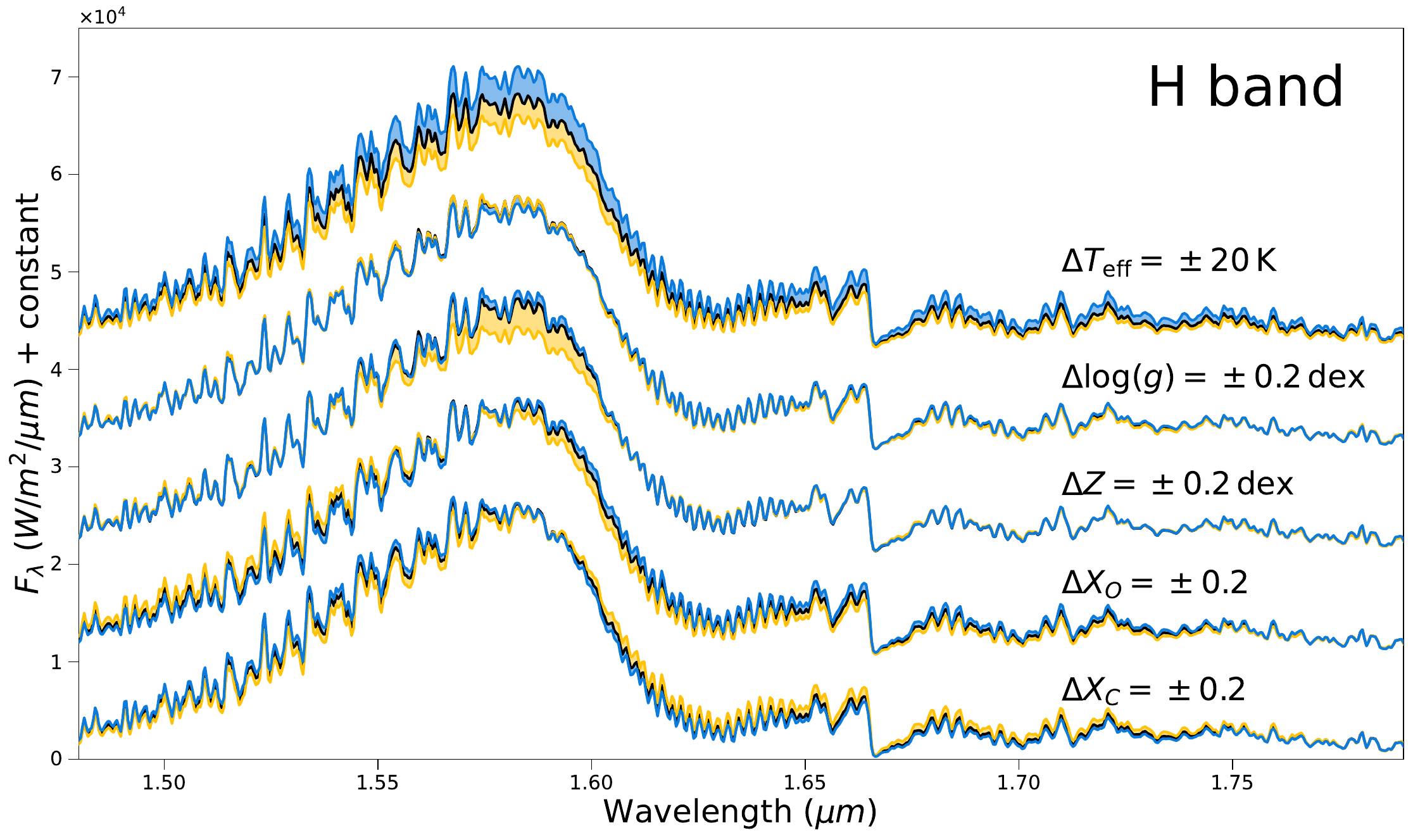}
\end{center}
\caption{Same as Figure \ref{fig:Yband_models} but for the $H$ band.\label{fig:Hband_models}}
\end{figure}

\begin{figure}[ht!]
\begin{center} 
\plotone{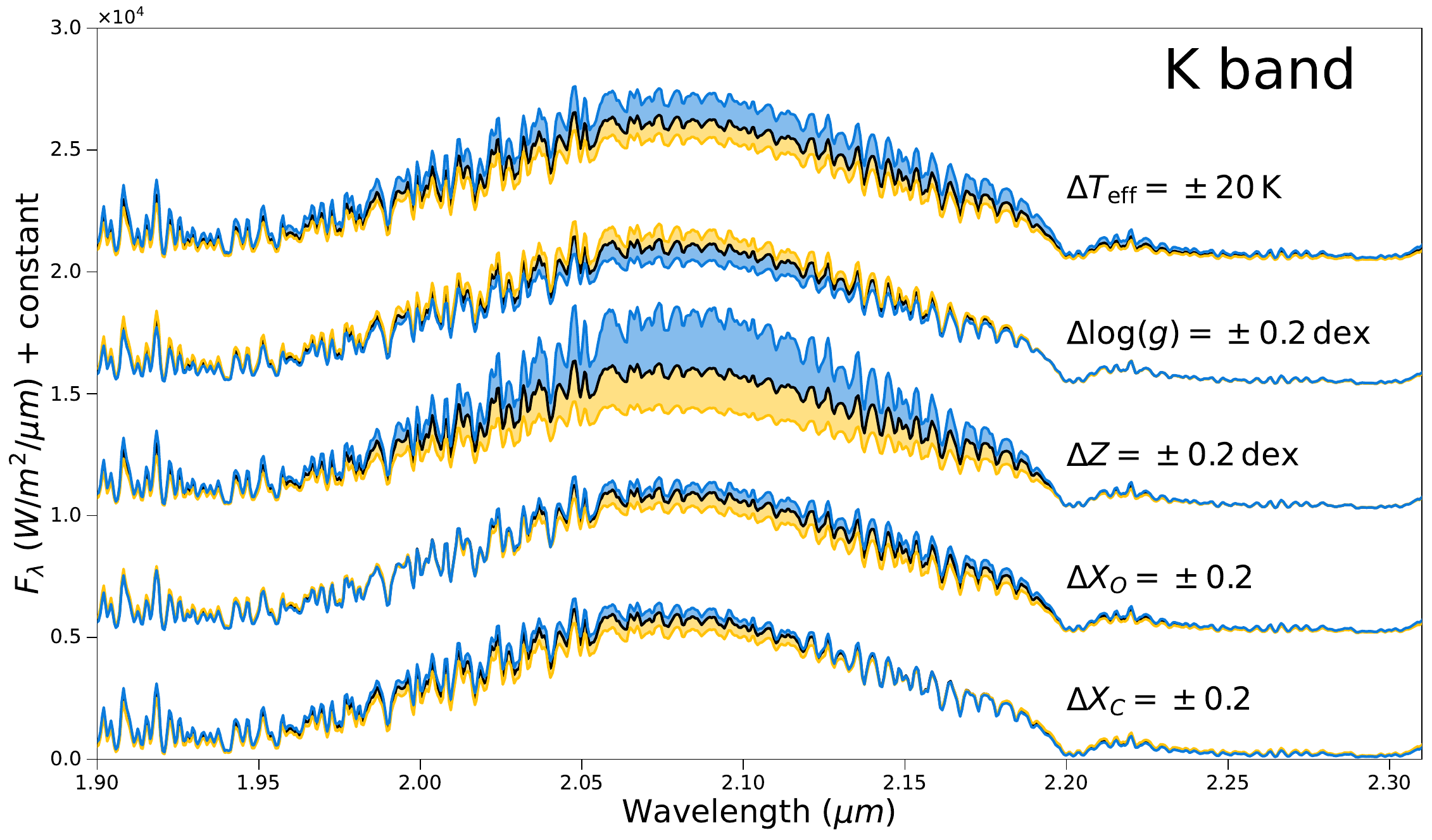}
\end{center}
\caption{Same as Figure \ref{fig:Yband_models} but for the $K$ band.\label{fig:Kband_models}}
\end{figure}

In this paper we have expanded the \texttt{ATMO} 2020 model atmospheres \citep{Phillips_2020} to include non-solar abundances in the mid/late-T dwarf regime ($T_\mathrm{eff}=700-900\,$K). This is done through altering the bulk metallicity of the atmosphere and the C/O ratio by separately changing the carbon and oxygen elemental abundances. We use these new models to assess the spectral signatures of non-solar abundances and C/O ratios at the resolution of the Gemini/GNIRS instrument in the $0.8-2.5\,\mu$m wavelength range (Figures \ref{fig:Yband_models}-\ref{fig:Kband_models}). We find that altering the oxygen abundance has effects in the $Y$ and $K$ bands that are degenerate with metallicity, but can be distinguished by changes in flux on the red side of the $H$ band. Altering the carbon abundance impacts the shape of the $H$ band and the flux on the blue side of the $K$ band. Our results indicate that non-solar abundances of oxygen and carbon could be distinguished from one another at this resolution based on the shapes of the $H$ and $K$ bands. 

\begin{figure}[ht!]
\plotone{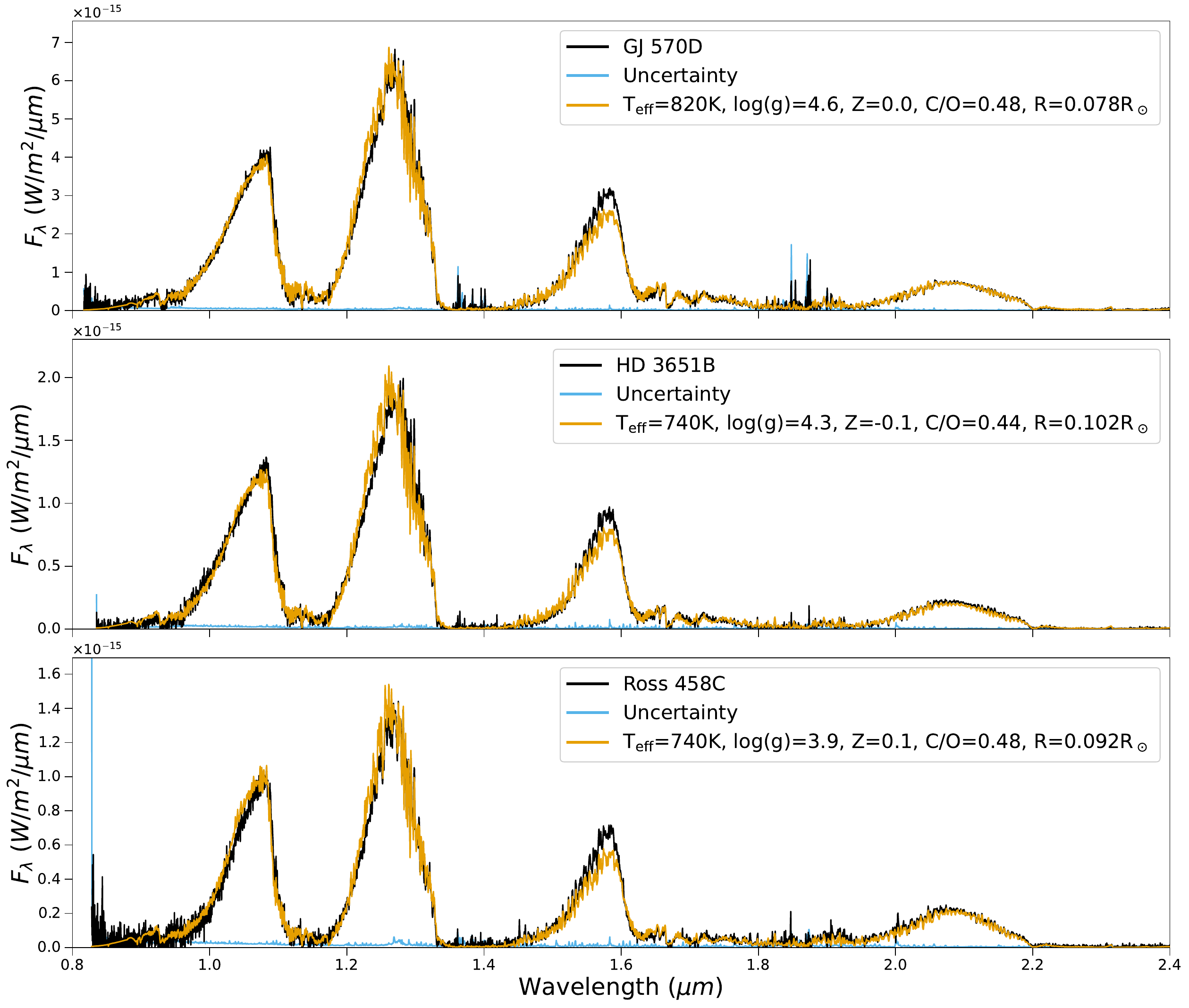}
\caption{ The best-fit \texttt{ATMO} spectra (\textit{orange}) to the GNIRS spectra (\textit{black}) of GJ 570D (top), HD 3651B (middle) and Ross 458C (bottom) . The uncertainties on the GNIRS spectra are plotted as the blue line at the bottom of each plot.  \label{fig:ALL_model_fit_fullspec}}
\end{figure}

\begin{figure}[ht!]
\plotone{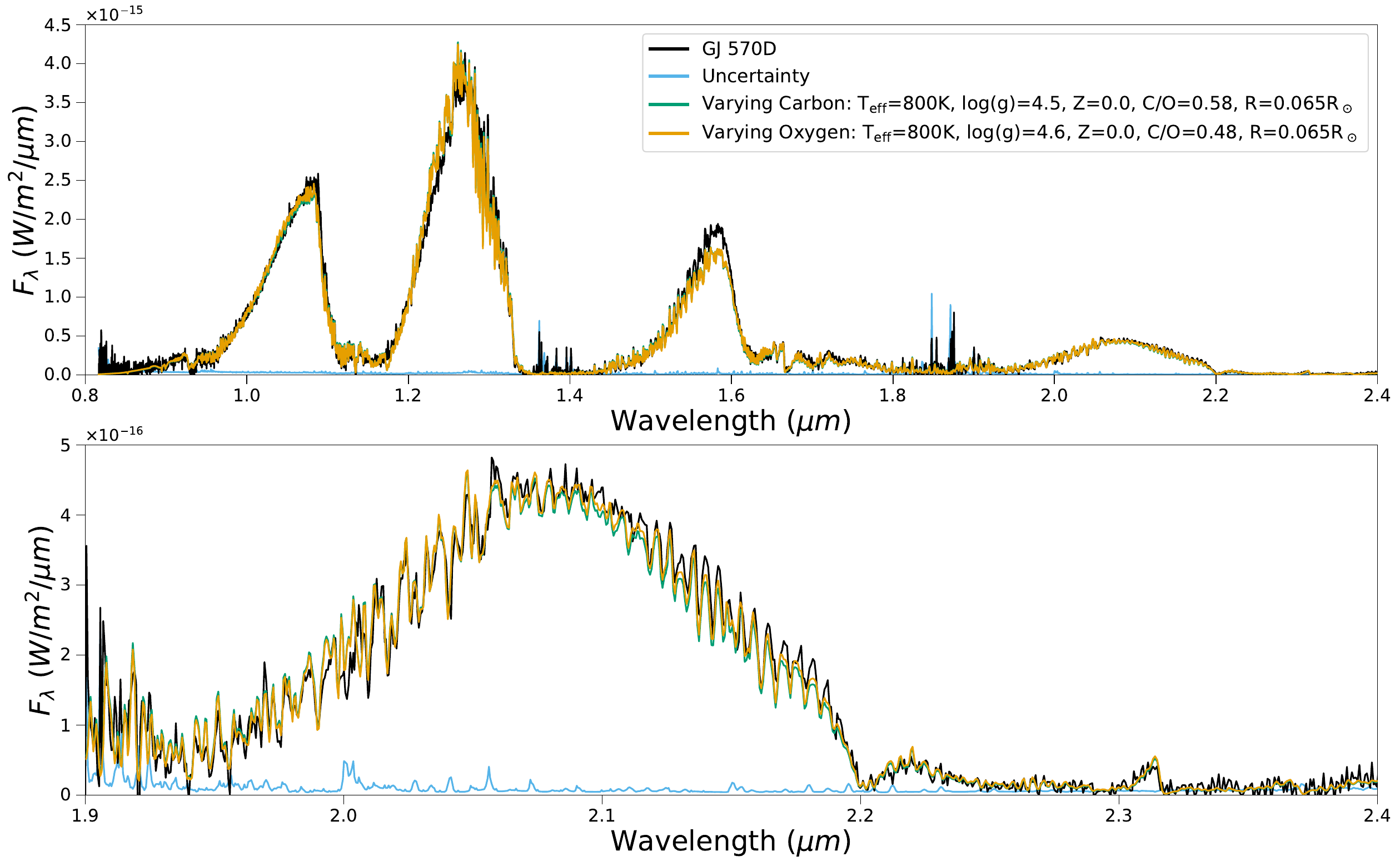}
\caption{The best-fit \texttt{ATMO} spectra to the GNIRS spectrum of GJ 570D when varying carbon (\textit{green}) and oxygen (\textit{orange}) in the model atmosphere. The uncertainties on the GNIRS spectrum are plotted as the blue line. Note that these model fits are limited to $\log(g)\ge4.5$, hence why the best-fit parameters when varying oxygen differ to those presented elsewhere in the paper (e.g. Figure \ref{fig:ALL_model_fit_fullspec}). \label{fig:varO_varC_comparison}}
\end{figure}

We compare our models with new Gemini/GNIRS $R\approx1700$ spectra of three benchmark T dwarfs, GJ 570D, HD 3651B and Ross 458C. Our model fitting analysis finds negligible differences in the best-fitting model parameters when changing the C/O ratio through either the carbon or oxygen abundances (Figure \ref{fig:varO_varC_comparison}), and we therefore use only models which alter the oxygen abundance of the model atmosphere. We find approximately solar C/O ratios for all objects and best-fitting parameters ($T_\mathrm{eff}$, $\log(g)$, $Z$) broadly consistent with other analyses in the literature (Tables \ref{tab:GJ_570D_params}, \ref{tab:HD3651B_params} and \ref{tab:ROSS_458C_params}, Figure \ref{fig:ALL_model_fit_fullspec}). 

\begin{figure}[ht!]
\plotone{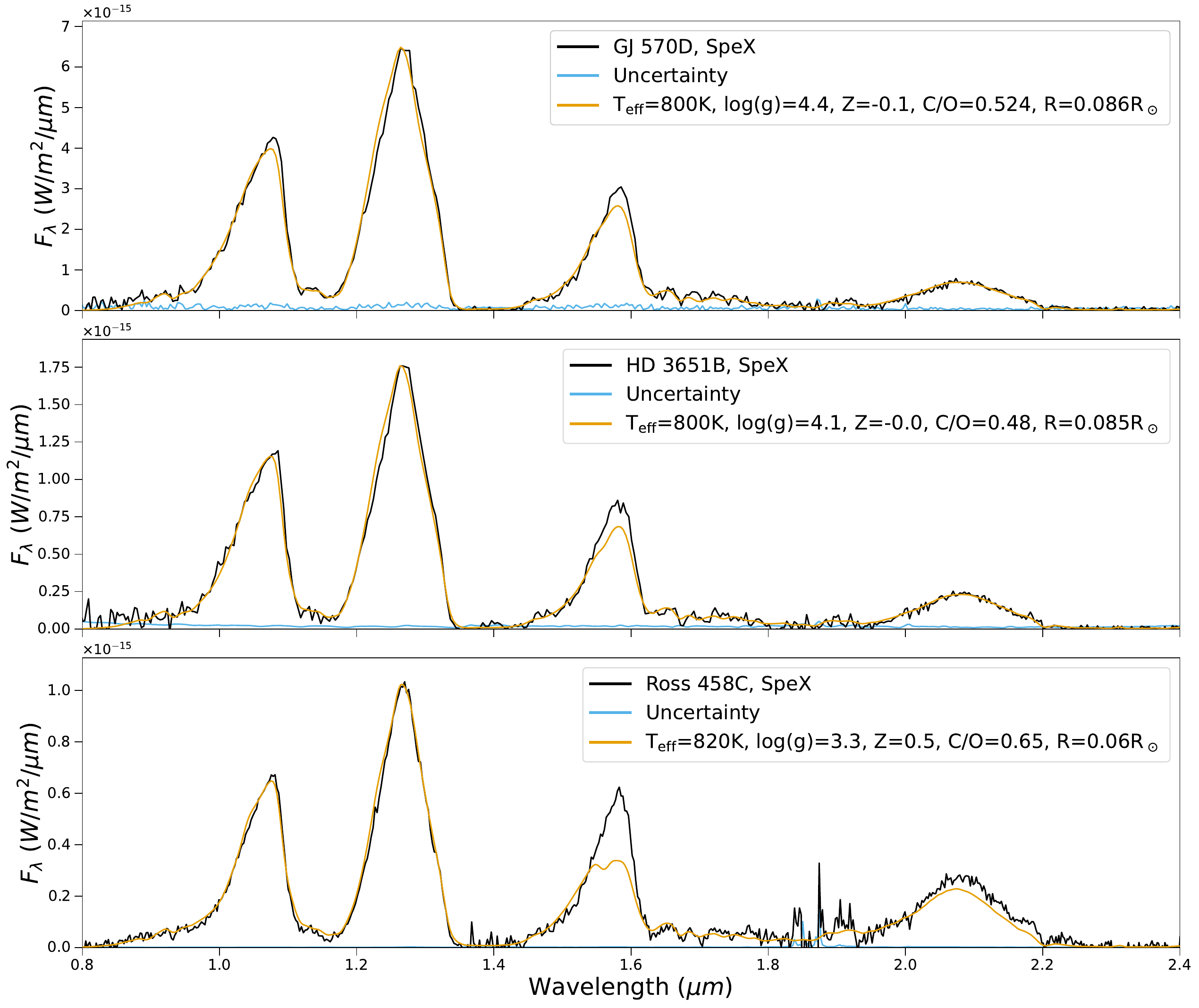}
\caption{ The best-fit \texttt{ATMO} spectrum (\textit{orange}) to the IRTF/SpeX spectra (\textit{black}) of GJ 570D (top) \citep{Burgasser_2004}, HD 3651B (middle) \citep{Burgasser_2007}, and Ross 458C (bottom) \citep{Zhang_2021_a}. The uncertainties on the SpeX spectra are plotted as the blue line in all panels. \label{fig:ALLSpeX_model_fit}}
\end{figure}

\begin{figure}[ht!]
\plotone{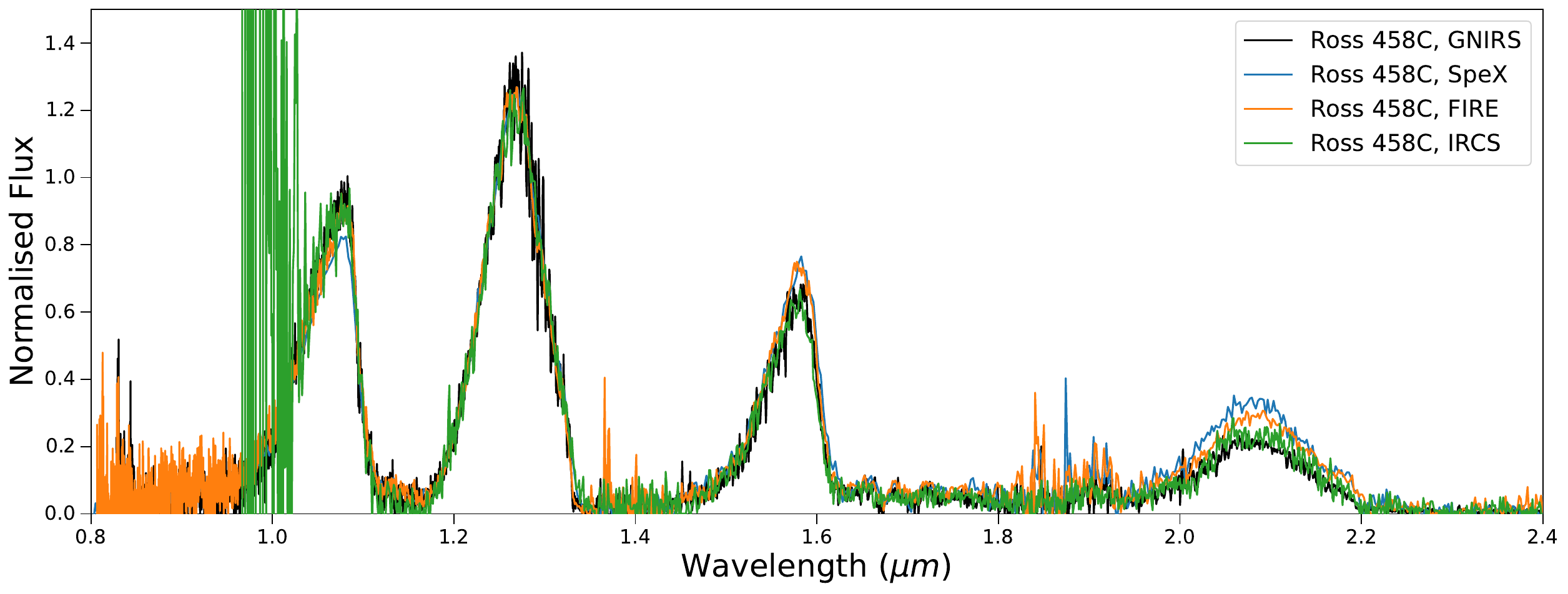}
\caption{Comparison of the Gemini/GNIRS (this work), IRTF/SpeX \citep{Zhang_2021_a}, Magellan/FIRE \citep{Burgasser_2010} and Subaru/IRCS \citep{Burningham_2011} spectra of Ross 458C. All spectra have been normalized to the median flux in the peak of the $J$ band ($1.225-1.300\,\mu$m). \label{fig:GNIRS-SpeX_Ross458C_comp}}
\end{figure}

To examine the benefits that $R\approx1700$ spectra provide over the more widely available low-resolution ($R\sim150$) IRTF/SpeX spectra, we also fit the \texttt{ATMO} models to the SpeX spectra of GJ 570D, HD 3651B and Ross 458C (Figure \ref{fig:ALLSpeX_model_fit}). For GJ 570D, we obtain the same gravity and C/O ratio as the best-fitting model to the higher-resolution GNIRS data. However, the best-fit model to the SpeX data gives a $\approx50\,$K cooler effective temperature and a somewhat sub-solar metallicity ($-0.1\,$dex). The higher-resolution GNIRS data appears to better establish the metallicity of this object, given that the metallicity is constrained to be approximately solar from observations of the host star.  Fitting the SpeX spectrum of HD 3651B we obtain the same gravity but a larger effective temperature compared to the best-fit of the GNIRS data, more similar to previous grid fitting analyses. The best-fit metallicity increases by $0.1\,$dex, but is still low compared to the metallicity of the host star.  Finally, we also fit the \texttt{ATMO} models to the SpeX spectrum of Ross 458C. However this analysis is complicated by the fact that the GNIRS and SpeX spectra differ substantially, particularly in the $J-H$ and $J-K$ colors. Further investigation reveals that the near-infrared ($0.8-2.5\,\mu$m) spectra of Ross 458C differs across different epochs taken with multiple spectrographs, indicating that this object could be potentially variable (Figure \ref{fig:GNIRS-SpeX_Ross458C_comp}).

\begin{figure}[ht!]
\plotone{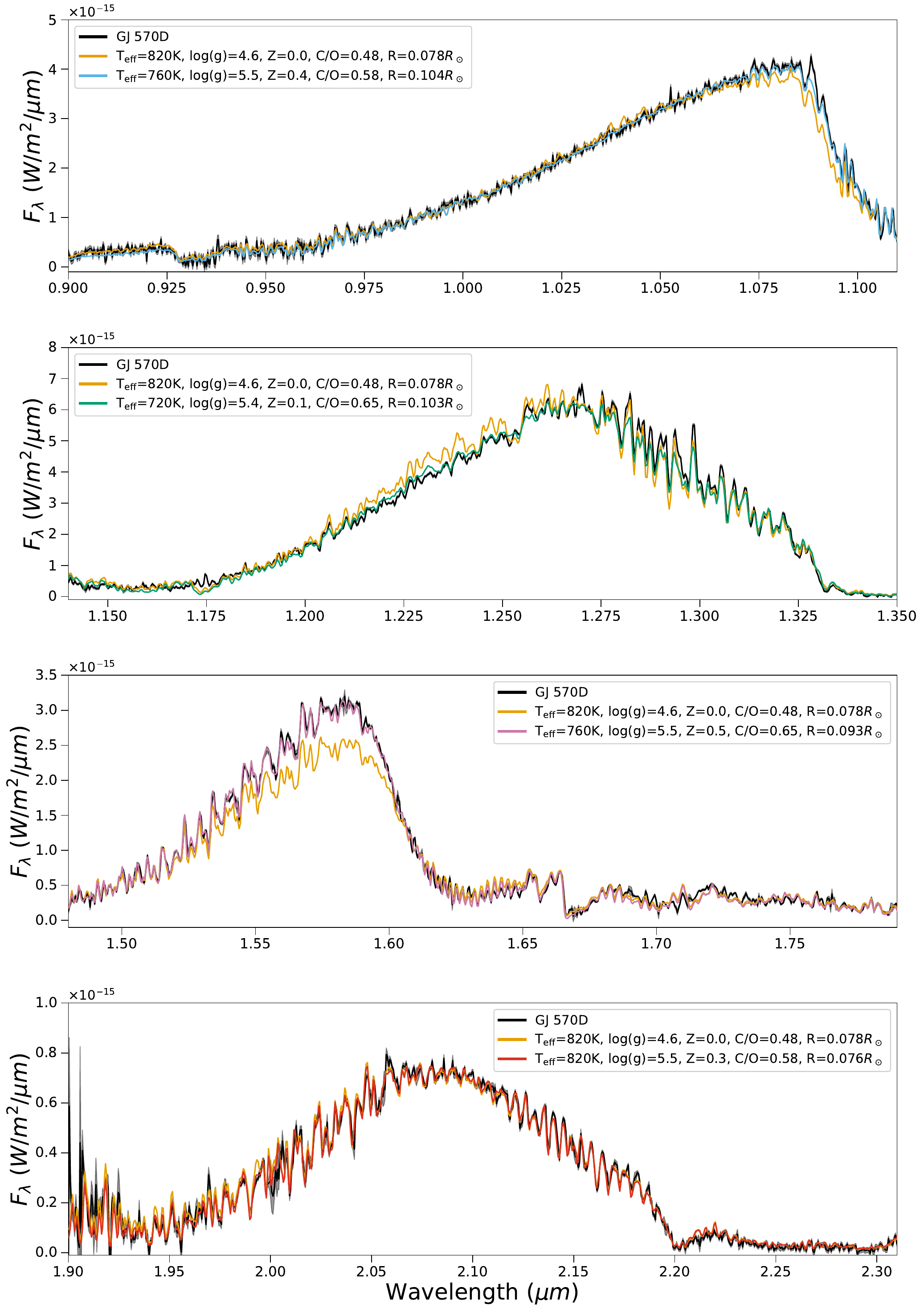}
\caption{$Y$-, $J$-, $H$-, and $K$-band (from top to bottom) spectra of GJ 570D including the best-fitting model to the full wavelength range (orange, same as Figure \ref{fig:ALL_model_fit_fullspec} model) and the best-fitting model to the individual wavelength ranges shown in each panel. \label{fig:GJ_570D_model_fit_zoomspec}}
\end{figure}

\begin{figure}[ht!]
\plotone{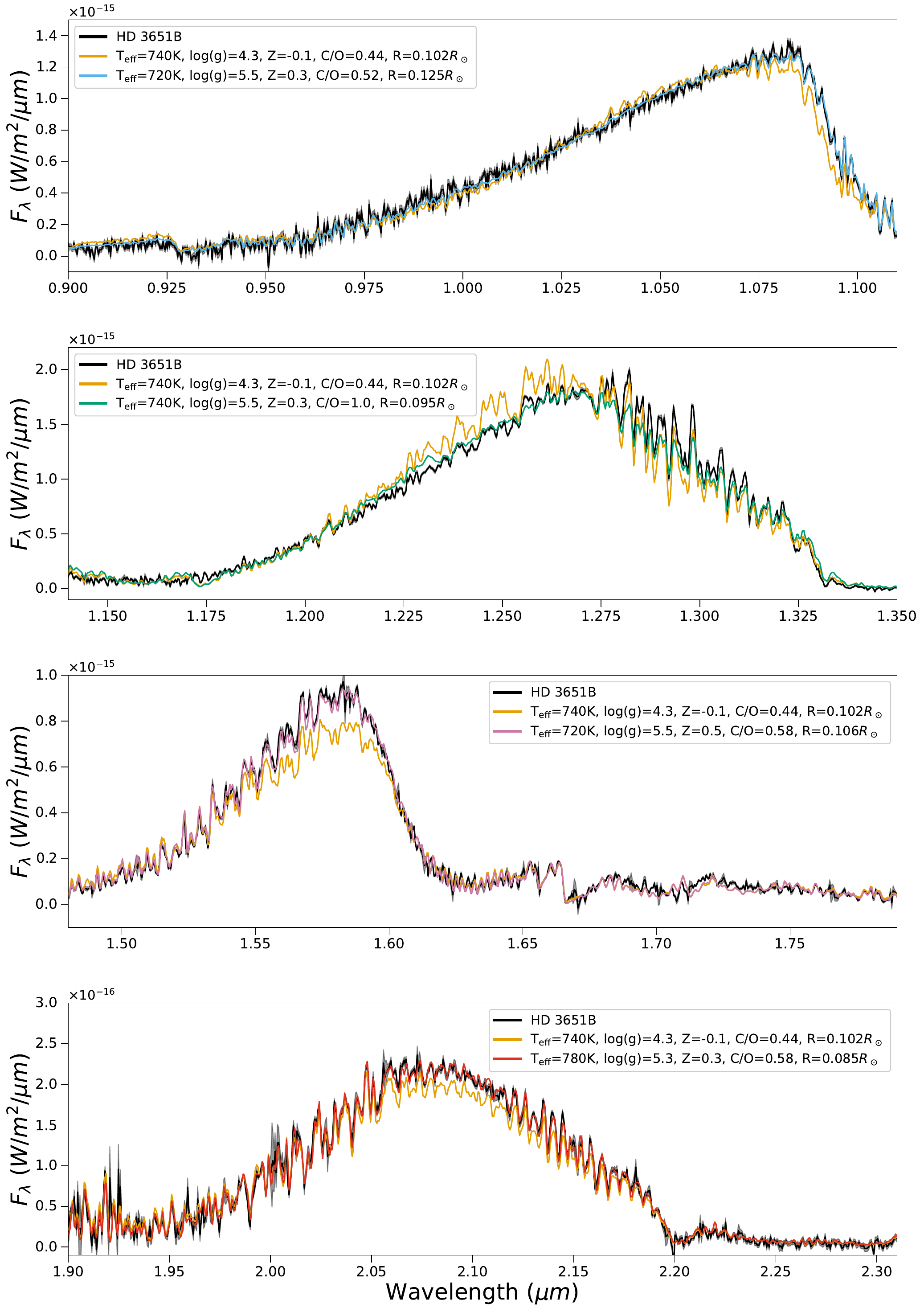}
\caption{ $Y$-, $J$-, $H$-, and $K$-band (from top to bottom) spectra of HD 3651B including the best-fitting model to the full wavelength range (orange, same as Figure \ref{fig:ALL_model_fit_fullspec}) and the best-fitting model to the individual wavelength ranges shown in each panel. \label{fig:HD3651B_model_fit_zoomspec}}
\end{figure}

\begin{figure}[ht!]
\plotone{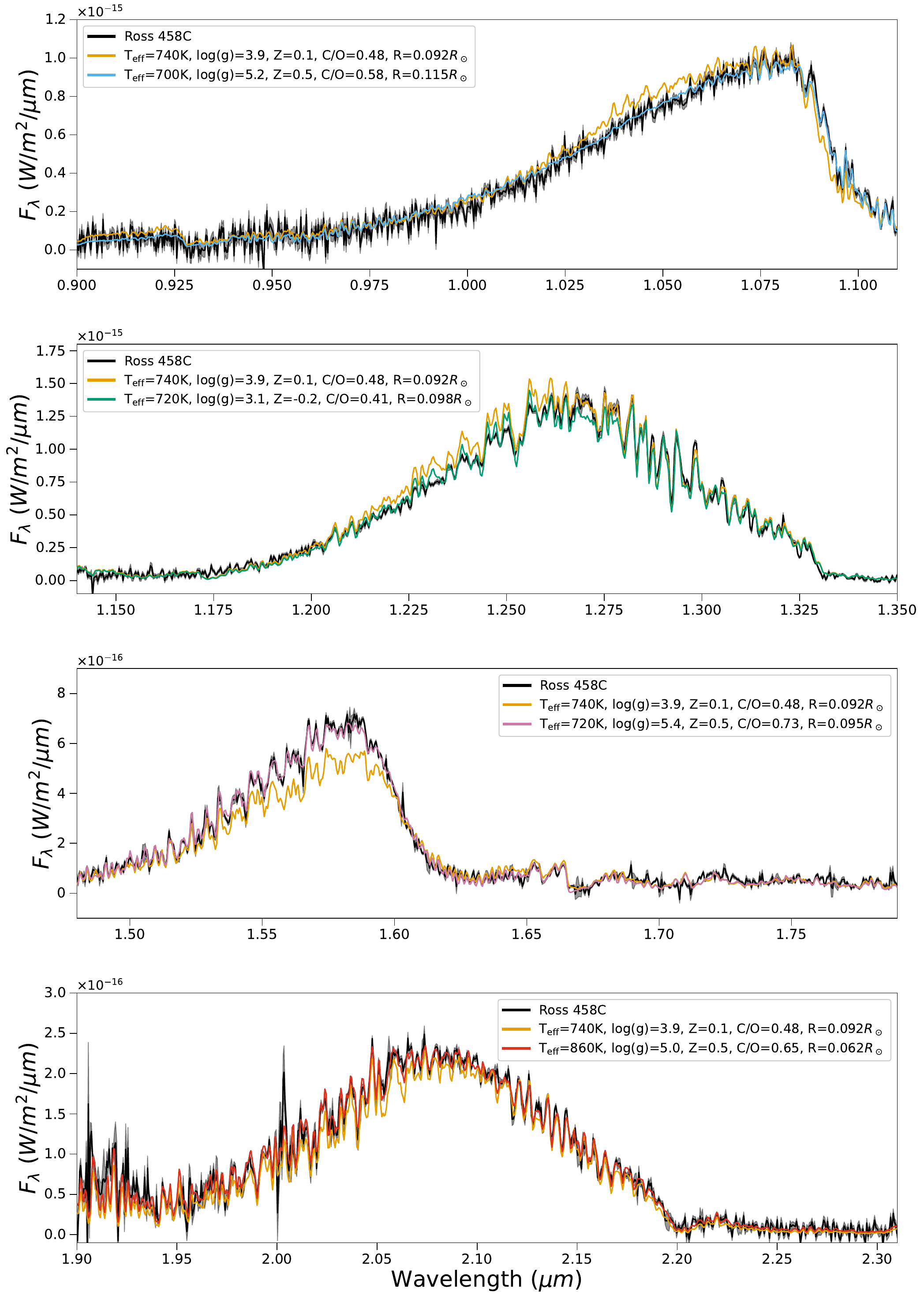}
\caption{$Y$-, $J$-, $H$-, and $K$-band (from top to bottom) spectra of Ross 458C including the best-fitting model to the full wavelength range (orange, same model as Figure \ref{fig:ALL_model_fit_fullspec}) and the best-fitting model to the individual wavelength ranges shown in each panel. \label{fig:ROSS_458C_model_fit_zoomspec}}
\end{figure}

\begin{figure}[ht!]
\plotone{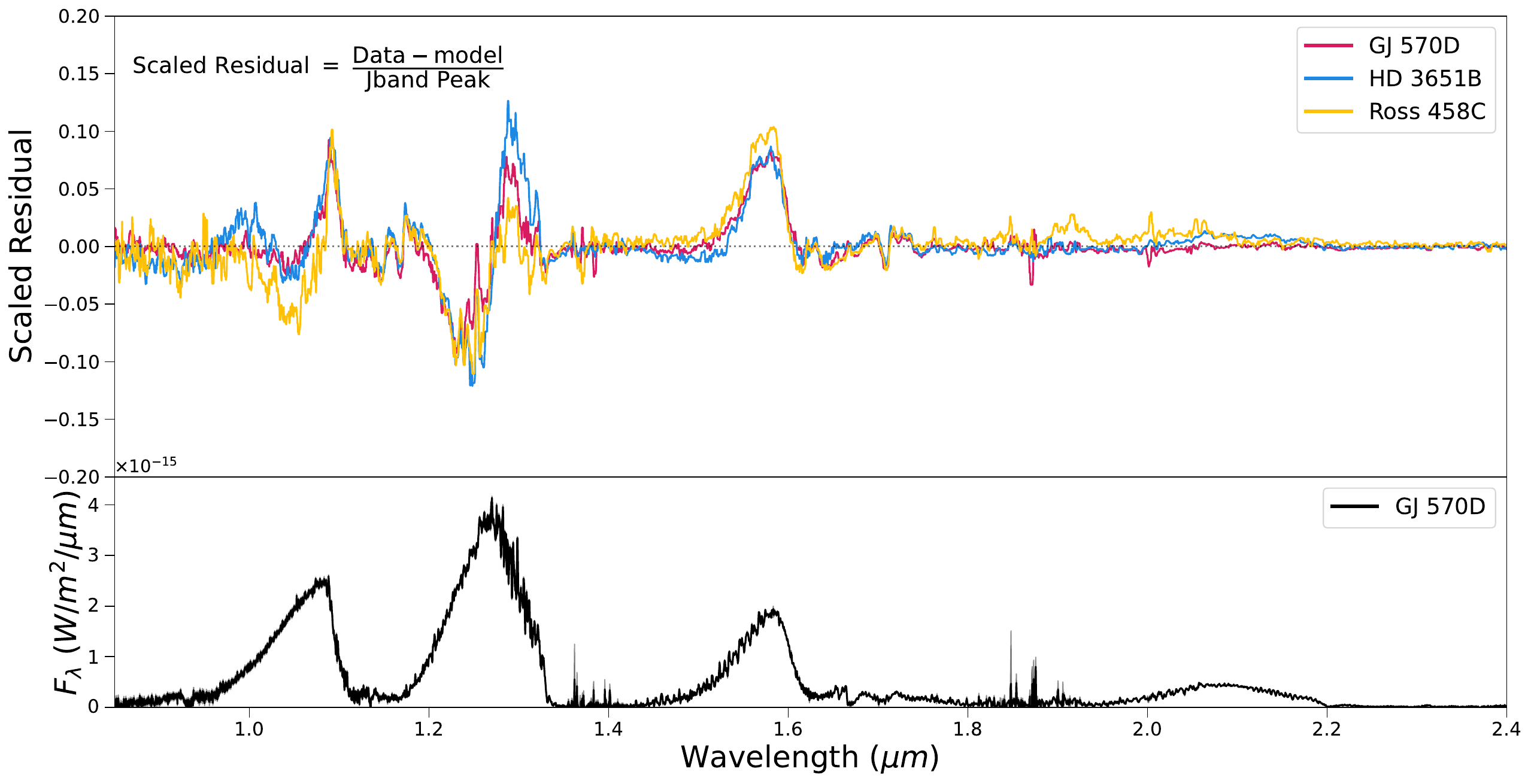}
\caption{Top: Spectral-fitting residuals for the 3 benchmark brown dwarfs analysed in this work. Each residual has been normalised by the object's median flux at the peak of the $J$-band. The residuals have been smoothed using a median filter with a window size of 10 pixels. Bottom: The GNIRS spectrum of GJ 570D as a reference.  \label{fig:benchmark_residuals}}
\end{figure}

There are three regions of the near-infrared spectrum that are not fitted satisfactorily, and these regions are consistent in the best fits of all objects. The best-fit models under-predict the flux on the red side ($\approx1.07-1.10\,\mu$m) of the $Y$ band,  overpredict and under predict the flux on the blue ($\approx1.22-1.26\,\mu$m) and red ($\approx1.27-1.31\,\mu$m) side of the $J$ band respectively,  and under-predict the peak flux ($\approx1.54-1.60\,\mu$m) of the $H$ band.  These discrepancies can be seen in the spectral residuals shown in Figure \ref{fig:benchmark_residuals}.  To investigate these discrepancies, we also fit the \texttt{ATMO} models to the $Y$, $J$, $H$ and $K$ bands individually (Figures \ref{fig:GJ_570D_model_fit_zoomspec}, \ref{fig:HD3651B_model_fit_zoomspec} and \ref{fig:ROSS_458C_model_fit_zoomspec}). These individual-band fits reproduce the flux and shape of each band significantly better than the full-spectrum fit. However, the resulting best-fit parameters are not consistent with each other and disagree with the results from the full spectrum. This leads us to conclude that the opacities used in the atmosphere model are sufficiently complete and accurate at this spectral resolution, and deficiencies are instead more likely to lie in the temperature profile and/or the corresponding chemical abundances. 

\begin{figure}[ht!]
\plotone{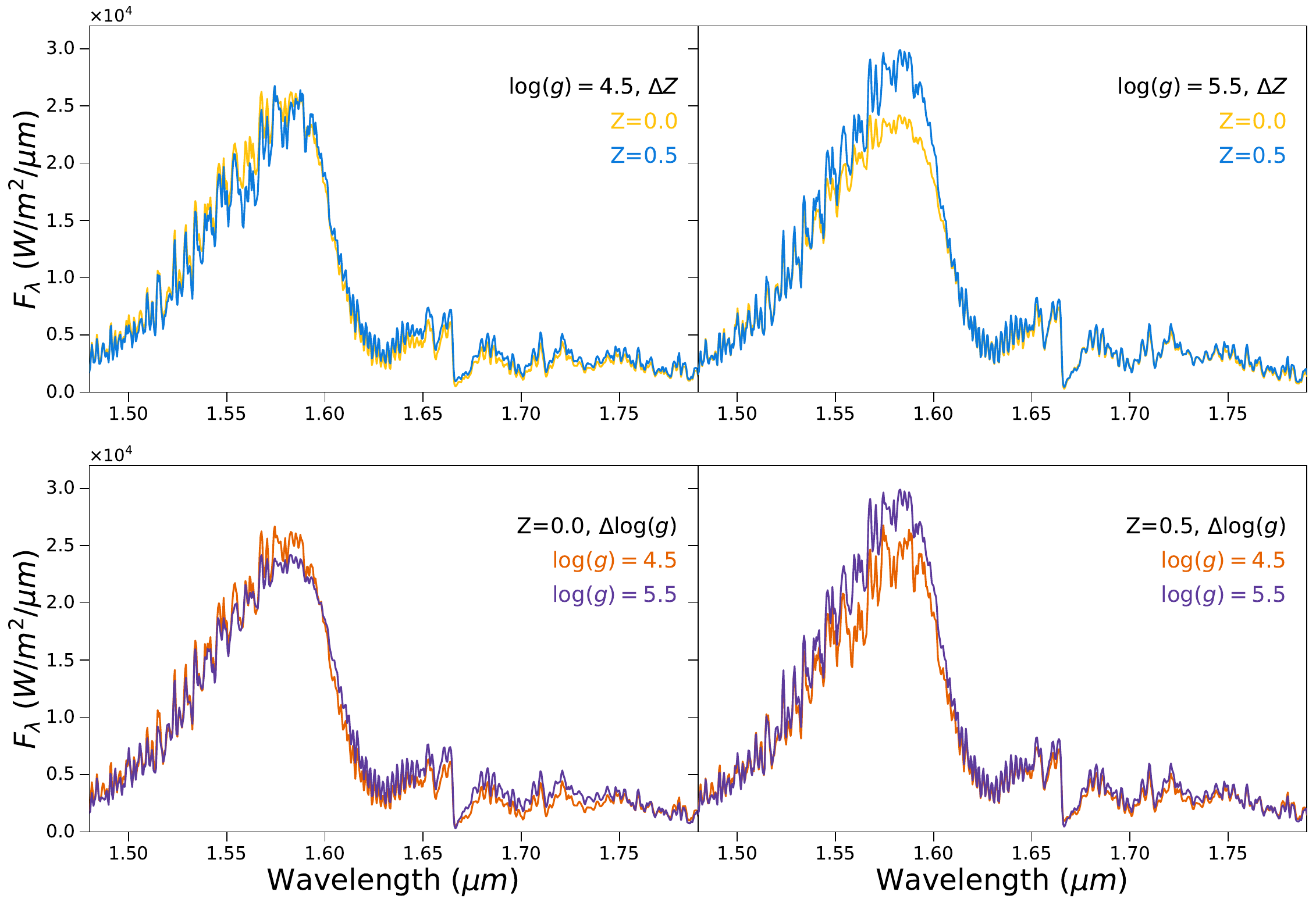}
\caption{Model H-band emission spectra for a $T_\mathrm{eff}=800\,$K atmosphere, and varying gravity and metallicity as indicated in the legend. Changing gravity/metallicity has a larger effect on the peak H-band flux at high metallicity/gravity. \label{fig:Hband_model_spec_analysis}}
\end{figure}

\begin{figure}[ht!]
\plotone{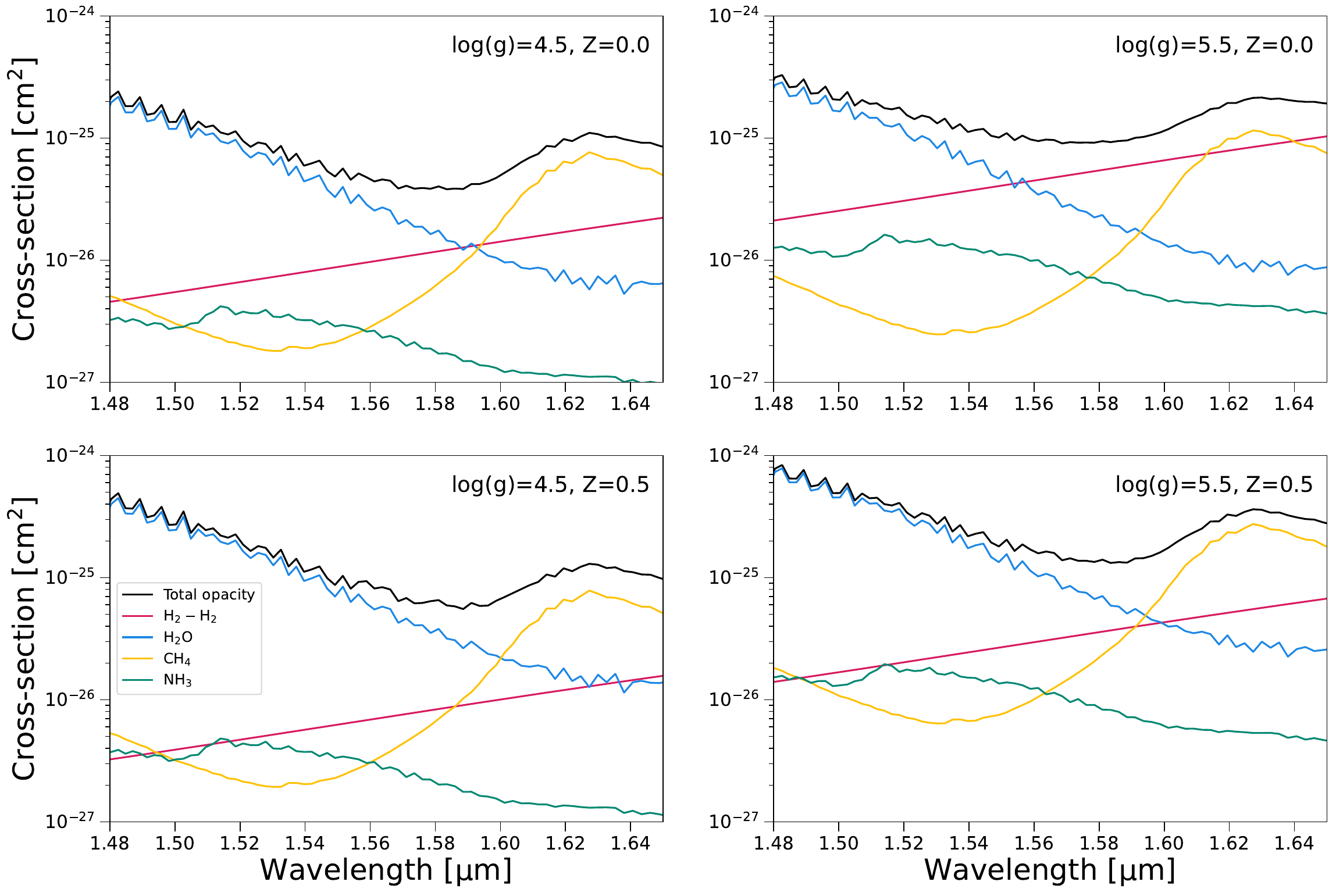}
\caption{H-band absorption cross-sections at the location of the peak of the contribution function at $1.58\,\mu$m, for a $T_\mathrm{eff}=800\,$K model atmosphere, with varying gravity and metallicity. The total mixture opacity (black) is the sum of the individual opacity sources (colored lines).   \label{fig:Hband_model_kabs_analysis}}
\end{figure}

\begin{figure}[ht!]
\plotone{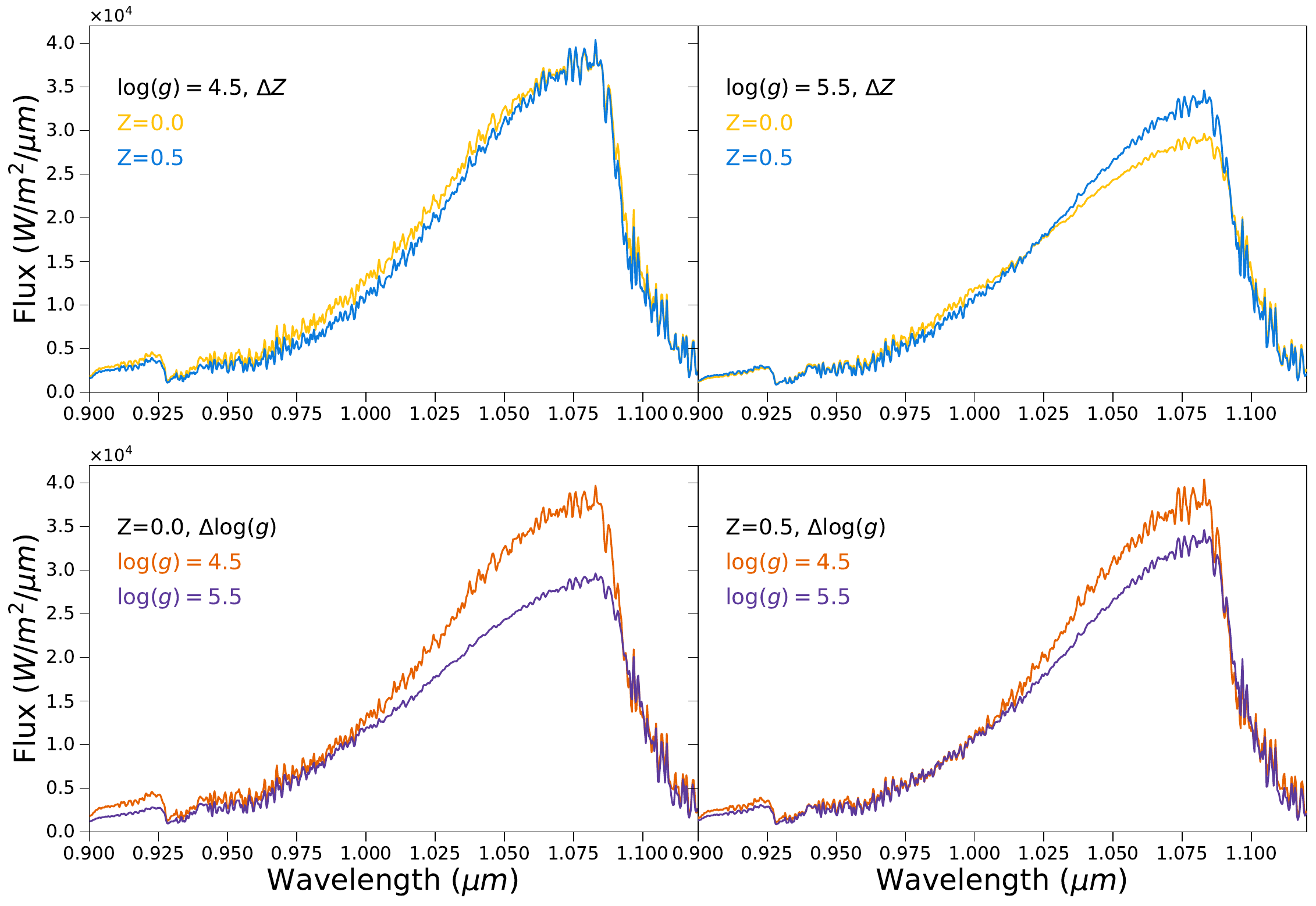}
\caption{Same as Figure \ref{fig:Hband_model_spec_analysis} but for the $Y$ band. \label{fig:Yband_model_spec_analysis}}
\end{figure}

\begin{figure}[ht!]
\plotone{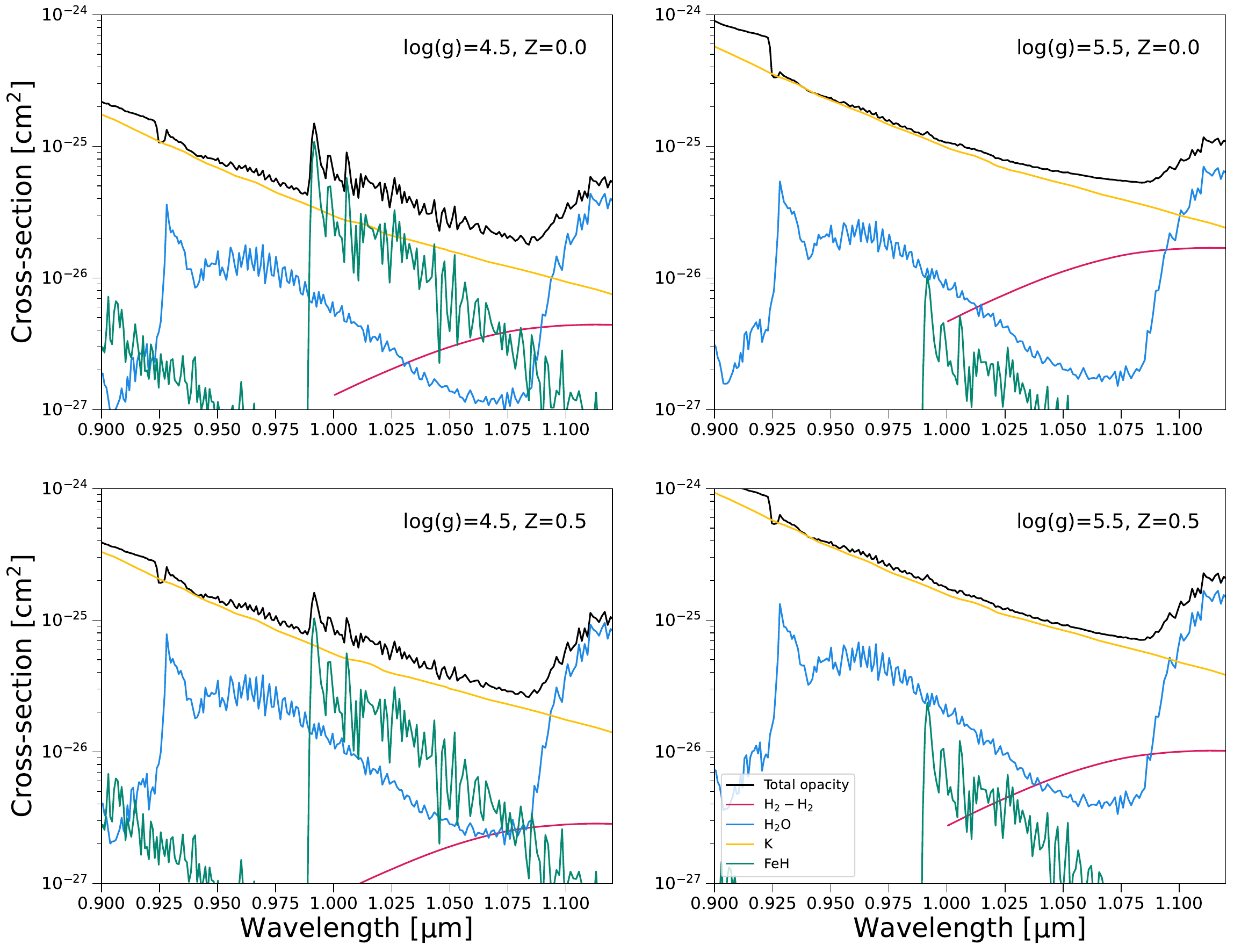}
\caption{Same as Figure \ref{fig:Hband_model_kabs_analysis} but showing absorption cross-sections at the location of the peak of the contribution function at $1.09\,\mu$m. \label{fig:Yband_model_kabs_analysis}}
\end{figure}

\begin{figure}[ht!]
\plotone{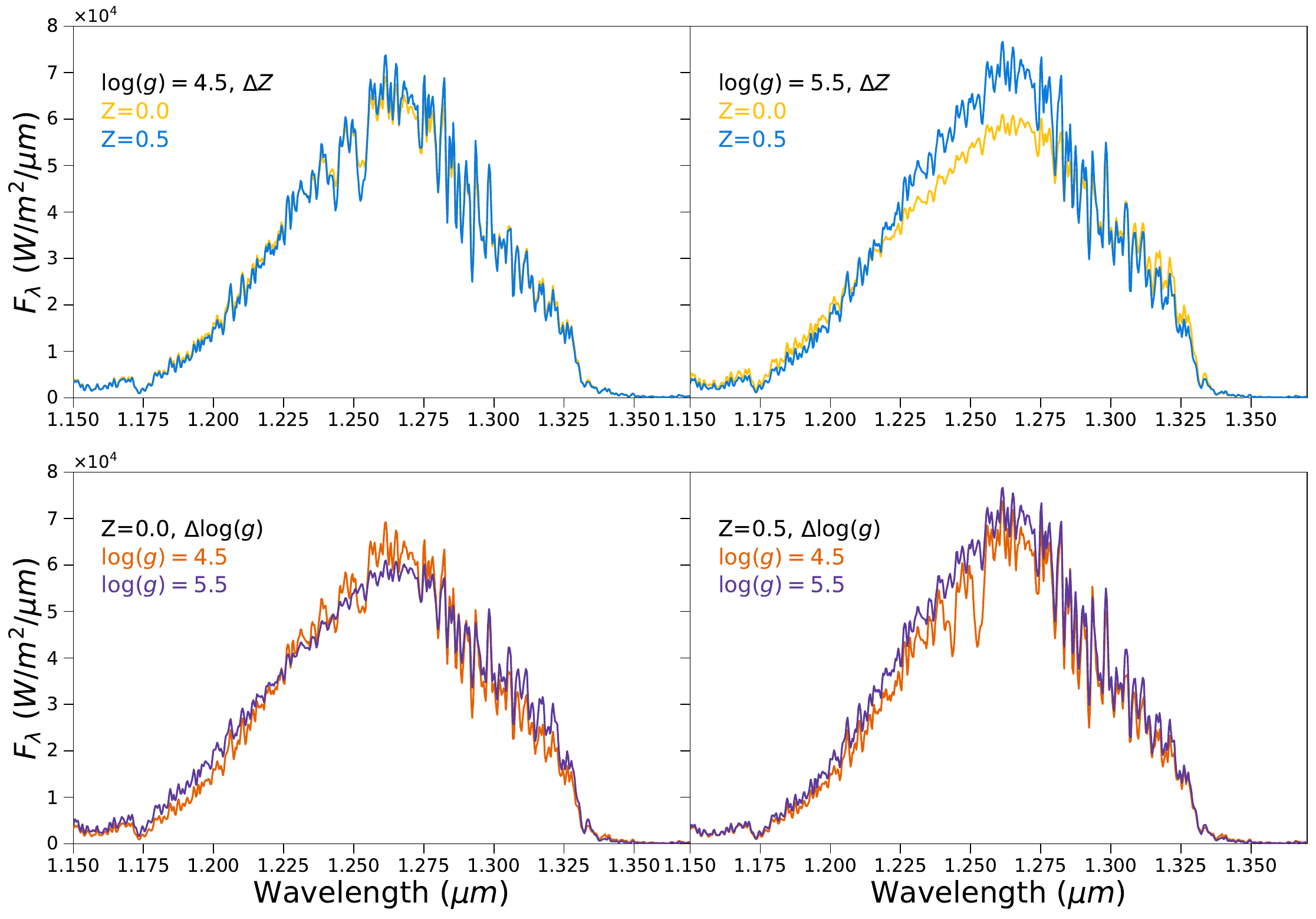}
\caption{Same as Figure \ref{fig:Hband_model_spec_analysis} but for the $J$ band. \label{fig:Jband_model_spec_analysis}}
\end{figure}

\begin{figure}[ht!]
\plotone{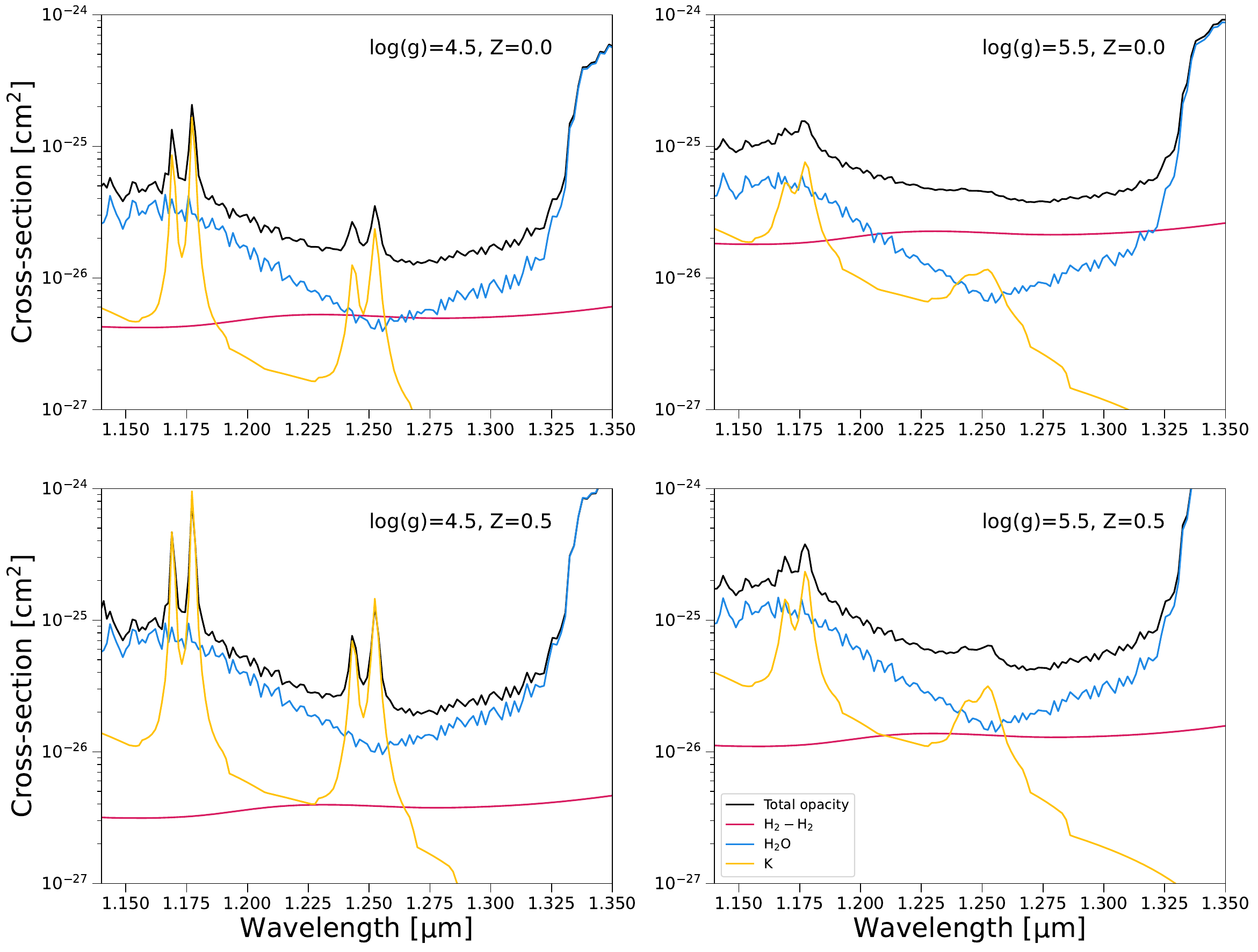}
\caption{Same as Figure \ref{fig:Hband_model_kabs_analysis} but showing absorption cross-sections at the location of the peak of the contribution function at $1.25\,\mu$m. \label{fig:Jband_model_kabs_analysis}}
\end{figure}

 While the individual band best-fit parameters are inconsistent with the results from the full spectrum, we find a consistent shift to high-gravity, high-metallicity atmospheres. This is particularly apparent in the individual $H$-band fits, which show a clear trend by moving to high gravity and high metallicity to better reproduce the $H$-band spectra of all objects.  By examining the model atmosphere properties at the $H$-band photosphere, we find that this is due to the interplay/competition of the effects of gravity and metallicity on $\mathrm{H_2-H_2}$ CIA (Figures \ref{fig:Hband_model_spec_analysis} and \ref{fig:Hband_model_kabs_analysis}). These parameters allow the model to control the peak $H$-band flux through the level of $\mathrm{H_2-H_2}$ CIA and drive the individual-band fits to converge on high-gravity, high-metallicity models.  This same process occurs in the individual-band fits of the $Y$ and $J$ bands, for which the best-fit parameters again shift to high gravity and high metallicity (except for the $J$-band fit of Ross 458C), with $\mathrm{H_2-H_2}$ CIA playing a key role (see Figures \ref{fig:Hband_model_kabs_analysis}, \ref{fig:Yband_model_kabs_analysis} and \ref{fig:Jband_model_kabs_analysis}).  Such high-gravity, high-metallicity atmospheres are unrealistic for all objects given the age and metallicity determined from their host stars, and thus it is clear that these discrepancies are due to incorrect or missing physics in the atmosphere models. 

\begin{figure}[ht!]
\plotone{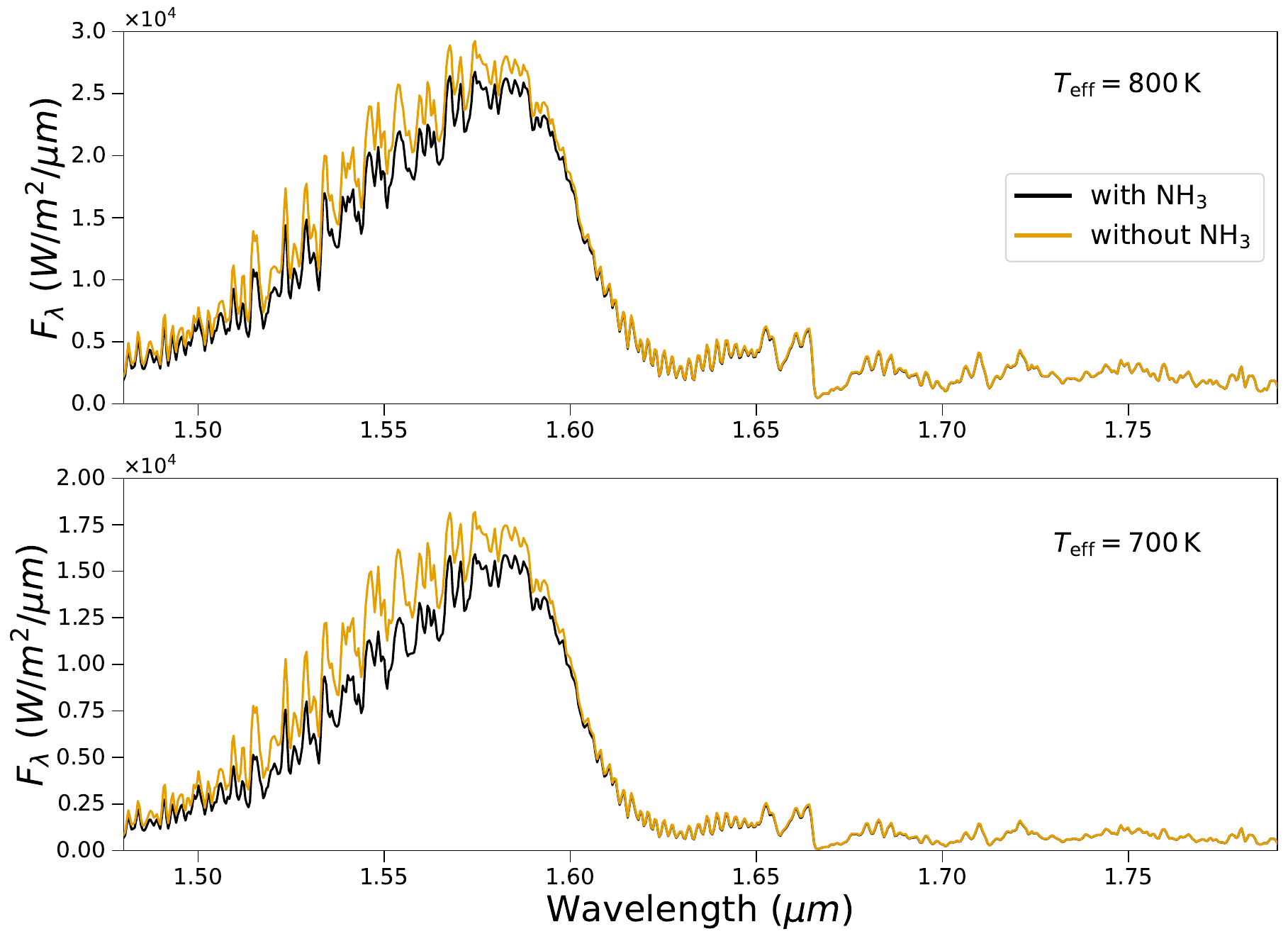}
\caption{Model emission spectra calculated with (black) and without (orange) $\mathrm{NH_3}$ opacity, for $T_\mathrm{eff}=800\,$K (top) and $700\,$K (bottom), $\log(g)=4.5$, $Z=0.0$ and $\mathrm{C/O}=0.55$ \label{fig:noNH3_comparison}}
\end{figure}

To explore the underprediction of the $H$-band further, we removed the $\mathrm{NH_3}$ opacity from the calculation of the model emission spectra, to simulate a more extreme quenching of $\mathrm{NH_3}$ due to vertical mixing than currently modeled by our non-equilibrium chemistry scheme. The quenching of $\mathrm{NH_3}$ has been shown to be particularly important in reproducing the $H$-band flux in observations of late-T and Y dwarfs \citep{Tremblin_2015}. We find that while removing $\mathrm{NH_3}$ opacity does brighten the $H$ band, it does not fully account for the under-prediction of flux seen in our best-fits of GJ 570D and Ross 458C (Figure \ref{fig:noNH3_comparison}). A full comparison of the impact the choice of non-equilibrium chemistry scheme (e.g. relaxation scheme [\citealp{Tsai_2018}], kinetics network [\citealp{Venot_2012}], reduced kinetics network [\citealp{Venot_2019}]) has on the predicted $H$-band peak flux in cool brown dwarfs is needed to better understand the discrepancies seen between model and data. 

 The wide-seperation companions studied in this work provide crucial benchmarks thanks to the age and composition constraints derived from their host stars. In future works, we will compare our new non-solar \texttt{ATMO} models to a wider sample of high-quality, medium-resolution Gemini/GNIRS spectra of cool T type brown dwarfs. This wider sample will include free-floating brown dwarfs in the solar neighbourhood and confirmed members of nearby young ($<100\,$Myr) moving groups \citep{Zhang_2021_c}. This will produce measurements of the fundamental properties and C/O ratios for cool brown dwarfs, and allow us to compare the different formation pathways among companions, young moving group members and field brown dwarfs. 

\begin{acknowledgements}
This research was funded in part by the Gordon and Betty Moore Foundation through grant GBMF8550 and by the National Science Foundation through grant NSF AST 1838016. Z. Z. acknowledges support from the NASA Hubble Fellowship grant HST-HF2-51522.001-A awarded by the Space Telescope Science Institute, which is operated by the Association of Universities for Research in Astronomy, Inc., for NASA, under contract NAS5-26555. This work has benefitted from The UltracoolSheet at http://bit.ly/UltracoolSheet, maintained by Will Best, Trent Dupuy, Michael Liu, Aniket Sanghi, Rob Siverd, and Zhoujian Zhang, and developed from compilations by \citet{Dupuy_Liu_2012, Dupuy_Kraus_2013, Deacon_2014, Liu_2016, Best_2018, Best_2021, Schneider_2023, Sanghi_2023}. This work was enabled by observations made from the Gemini North telescope, located within the Maunakea Science Reserve and adjacent to the summit of Maunakea. We are grateful for the privilege of observing the Universe from a place that is unique in both its astronomical quality and its cultural significance.  
\end{acknowledgements}
\software{Astropy \citep{astropy_2013, astropy_2018, astropy_2022},
IPython \citep{Ipython_2007}, Numpy \citep{Numpy_2020}, Scipy \citep{Scipy_2020}, Matplotlib \citep{matplotlib_2007}.}

\bibliography{refs}{}
\bibliographystyle{aasjournal}

\newpage

\begin{rotatetable}
\movetableright=0.5mm
\begin{deluxetable*}{lccccccccc}
\tablenum{1}
\tablecaption{Gemini North/GNIRS spectroscopic observations. The tabulated $S/N$ values are the median in the wavelength ranges approximately corresponding to the $Y$ ($1.025-1.10\,\mu$m) $J$ ($1.20-1.35\,\mu$m), $H$ ($1.53-1.61\,\mu$m) and $K$ ($2.0-2.19\,\mu$m) peaks. \label{tab:gnirs_observations}}
\tablewidth{0pt}
\tablehead{\colhead{Target} & \colhead{$J_\mathrm{MKO}$} & \colhead{Date} & 
           \colhead{Exposure time} & \colhead{$S/N_{Y}$} & \colhead{$S/N_{J}$} & \colhead{$S/N_{H}$} &\colhead{$S/N_{K_s}$} & 
           \colhead{Airmass} & \colhead{Telluric standard} \\ 
            & (mag) & & (sec)}
\startdata
GJ 570D & 14.82 & 2021 Jul 9 & 756 & 66.7 & 101.6 & 88.5 & 48.9 & $1.70-1.78$ & HIP 73867\\ 
 HD 3651B & 16.16 & 2022 Aug 8 & 2240 & 56.2 & 86.0 & 70.3 & 37.1 & $1.13-1.23$ & HIP 6193\\ 
Ross 458C & 16.69 & 2021 Jul 1 & 3912 & 34.5 & 45.6 & 46.8 & 31.2 & $1.10-1.31$ & HIP 68868\\ 
%& & 2021 Jul 9 & 1304 & & & & & $1.56-1.72$
\enddata
\end{deluxetable*}
\end{rotatetable}

\begin{rotatetable}
\movetableright=0.1mm
%\movetabledown=5in
\begin{deluxetable*}{lccccccccc}
\tablenum{2}
\tablecaption{Parameters derived for GJ 570D from spectroscopic analysis \label{tab:GJ_570D_params}}
\tablewidth{0pt}
\tablehead{\colhead{$T_\mathrm{eff}$ (K)} & \colhead{$\log(g)$} & \colhead{$Z$} & 
           \colhead{C/O} & \colhead{Instrument} & \colhead{Analysis technique} & \colhead{Model grid} &\colhead{Reference} \\ 
    }
\startdata
$800^{+50}_{-50}\,$ & $4.6^{+0.25}_{-0.25}$ & $0.0^{+0.25}_{-0.25}$ & $0.48^{+0.23}_{-0.1}$ & GNIRS & $\chi^2$ minimisation & \texttt{ATMO} & This work \\
$800^{+50}_{-50}$ & $4.4^{+0.25}_{-0.25}$ & $-0.1^{+0.25}_{-0.25}$ & $0.52^{+0.23}_{-0.1}$ & SpeX & $\chi^2$ minimisation & \texttt{ATMO} & This work \\
$810^{+10}_{-10}$ & $5.16^{+0.07}_{-0.07}$ & $0.0$\tablenotemark{a} & 0.458\tablenotemark{a} & SpeX + mid IR & Luminosity analysis\tablenotemark{b} & SM\tablenotemark{c} & \citet{Saumon_2006} \\
%& & & & & atmosphere analysis & & & \\  
$808^{+43}_{-27}\,$ & $4.93^{+0.38}_{-0.55}$ & 0.0\tablenotemark{a} & 0.458\tablenotemark{a} & SpeX & Machine learning & \texttt{Sonora} & \citet{Oreshenko_2020} \\
$800^{+14}_{-100}\,$ & $5.08^{+0.62}_{-0.68}$ & 0.0\tablenotemark{a} & 0.55\tablenotemark{a} & SpeX & Machine learning & \texttt{HELIOS} & \citet{Oreshenko_2020} \\
$878^{+23}_{-78}\,$ & $5.27^{+0.43}_{-0.67}$ & 0.0\tablenotemark{a} & 0.55\tablenotemark{a} & SpeX & Machine learning & \texttt{AMES-Cond} & \citet{Oreshenko_2020} \\
$828^{+26}_{-25}\,$ & $3.90^{+0.25}_{-0.25}$ & $-0.33^{+0.14}_{-0.14}$ & 0.458\tablenotemark{a} & SpeX & Bayesian inference & \texttt{Sonora} & \citet{Zhang_2021_a}\\ 
$715^{+20}_{-22}\,$ & $4.8^{+0.3}_{-0.3}$ & $-0.15^{+0.07}_{-0.09}$ & $0.79^{+0.28}_{-0.23}$ & SpeX & Retrieval & - & \citet{Line_2015, Line_2017} \\
$730^{+18}_{-17}\,$ &  $4.61^{+0.08}_{-0.08}$ & $-0.15^{+0.05}_{-0.04}$ & $0.83^{+0.09}_{-0.8}$ & SpeX & Retrieval & - & \citet{Kitzmann_2020} \\
$789^{+4.05}_{-12.67}$ & $5.14^{+0.14}_{-0.26}$ & $-0.02^{+0.10}_{-0.10}$ & $0.77^{+0.08}_{-0.10}$ & SpeX & Retrieval & - & \citet{Zalesky_2022} \\
$722^{+23}_{-26}$ & $4.93^{+0.11}_{-0.12}$ & $-0.19^{+0.05}_{-0.03}$ & $0.87^{+0.08}_{-0.07}$ & SpeX & Retrieval & - & \citet{Whiteford_2023} \\
\enddata
\tablenotetext{a}{These values are fixed in the model fitting analysis.}
\tablenotetext{b}{Luminosity analysis refers to using spectroscopic and photometric observations, along with atmosphere models, to estimate the bolometric luminosity of an object. Combined with age estimates, the bolometric luminosity can be used to estimate a range of $T_\mathrm{eff}$ and $\log(g)$ from evolutionary tracks, as described in \citet{Saumon_2006}.}  
\tablenotetext{c}{SM refers to atmosphere models from the Saumon \& Marley group \citep{Saumon_2006, Saumon_2008, Saumon_2012, Morley_2012}.}
\end{deluxetable*}
\end{rotatetable}

%%% Table of previous work for HD 3651B
\begin{rotatetable}
\movetableright=0.5mm
\begin{deluxetable*}{lccccccccc}
\tablenum{3}
\tablecaption{ Parameters derived for HD 3651B from spectroscopic analysis \label{tab:HD3651B_params}}
\tablewidth{0pt}
\tablehead{\colhead{$T_\mathrm{eff}$ (K)} & \colhead{$\log(g)$} & \colhead{$Z$} & 
           \colhead{C/O} & \colhead{Instrument} & \colhead{Analysis technique} & \colhead{Model grid} &\colhead{Reference} \\ 
    }
\startdata
$740^{+50}_{-50}\,$ & $4.3^{+0.25}_{-0.25}$ & $-0.1^{+0.25}_{-0.25}$ & $0.44^{+0.23}_{-0.1}$ & GNIRS & $\chi^2$ minimisation & \texttt{ATMO} & This work \\
$800^{+50}_{-50}\,$ & $4.1^{+0.25}_{-0.25}$ & $0.0^{+0.25}_{-0.25}$ & $0.48^{+0.23}_{-0.1}$ & SpeX & $\chi^2$ minimisation & \texttt{ATMO} & This work \\
$818^{+28}_{-28}\,$ & $3.94^{+0.29}_{-0.28}$ & $-0.22^{+0.16}_{-0.16}$ & 0.458\tablenotemark{a} & SpeX & Bayesian inference & \texttt{Sonora} & \citet{Zhang_2021_a} \\
$719^{+19}_{-25}\,$ & $5.12^{+0.1}_{-0.2}$ & $0.08^{+0.05}_{-0.06}$ & $0.89^{+0.2}_{-0.17}$ & SpeX & Retrieval & - & \citet{Line_2015, Line_2017} \\
$715.57^{+24.68}_{-9.65}\,$ & $5.07^{+0.09}_{-0.14}$ & $-0.04^{+0.08}_{-0.07}$ & $0.98^{+0.11}_{-0.11}$ & SpeX & Retrieval & - & \citet{Zalesky_2022} \\
$790^{+30}_{-30}\,$ & $5.0^{+0.3}_{-0.3}$ & $+0.12^{+0.04}_{-0.04}$ & $0.55\tablenotemark{a}$ & SpeX & Semi-empirical & Tucson \tablenotemark{c} & \citet{Burgasser_2007} \\
$825^{+5}_{-5}\,$ & $5.45^{+0.05}_{-0.05}$ & $+0.2^{+0.1}_{-0.1}$ & $0.458\tablenotemark{a}$ & GNIRS & By-eye & SM\tablenotemark{d} & \citet{Leggett_2007} \\
$800$ & $4.5$ & $0.0\tablenotemark{a}$ & $0.55\tablenotemark{a}$ & SpeX & $\chi^2$ minimisation & \texttt{AMES-Cond} & \citet{Liu_2011} \\
$797$ & $4.8$ & $0.0\tablenotemark{a}$ & $0.55\tablenotemark{a}$ & SpeX & $\chi^2$ minimisation & Tucson \tablenotemark{c} & \citet{Liu_2011} \\
$900$ & $5.0$ & $0.0\tablenotemark{a}$ & $0.55\tablenotemark{a}$ & SpeX & $\chi^2$ minimisation & \texttt{BT-Settl} & \citet{Liu_2011} \\
\enddata
\tablenotetext{a}{These values are fixed in the model fitting analysis.} 
\tablenotetext{b}{A semi-empirical analysis comparing measured spectral indices to those from model spectra that have been calibrated on GJ 570D, as described in \citet{Burgasser_2006b} and \citet{Burgasser_2007}.}  
\tablenotetext{c}{Tucson refers to models from the Tucson group \citep{Burrows_2003, Burrows_2006}.} 
\tablenotetext{d}{SM refers to atmosphere models from the Saumon \& Marley group \citep{Saumon_2006, Saumon_2008, Saumon_2012, Morley_2012}.} 
\end{deluxetable*}
\end{rotatetable}

%%% Table of previous work for Ross 458C
\begin{rotatetable}
\movetableright=0.5mm
\begin{deluxetable*}{lccccccccc}
\tablenum{4}
\tablecaption{Parameters derived for Ross 458C from spectroscopic analysis \label{tab:ROSS_458C_params}}
\tablewidth{0pt}
\tablehead{\colhead{$T_\mathrm{eff}$ (K)} & \colhead{$\log(g)$} & \colhead{$Z$} & 
           \colhead{C/O} & \colhead{Instrument} & \colhead{Analysis technique} & \colhead{Model grid} &\colhead{Reference} \\ 
    }
\startdata
$740^{+50}_{-50}\,$ & $3.9^{+0.25}_{-0.25}$ & $0.1^{+0.25}_{-0.25}$ & $0.48^{+0.23}_{-0.1}$ & GNIRS & $\chi^2$ minimisation & \texttt{ATMO} & This work \\
$635^{+25}_{-35}$ & 4.0 & 0.0\tablenotemark{a} & 0.458\tablenotemark{a} & FIRE & $\chi^2$ minimisation & SM\tablenotemark{c} & \citet{Burgasser_2010} \\ 
$695^{+60}_{-60}$ & $4.0-4.7$ & 0.0\tablenotemark{a} & 0.55\tablenotemark{a} & IRCS & Luminosity analysis\tablenotemark{b} & SM\tablenotemark{c}+BT-Settl & \citet{Burningham_2011} \\  
$700$ & $4.0$ & 0.0\tablenotemark{a} & 0.458\tablenotemark{a} & FIRE & By-eye & SM\tablenotemark{c} & \citet{Morley_2012} \\
$804^{+30}_{-29}\,$ & $4.09^{+0.02}_{-0.01}$ & $0.50^{+0.00}_{-0.00}$ & 0.458\tablenotemark{a} & SpeX & Bayesian inference & \texttt{Sonora} & \citet{Zhang_2021_a} \\
$762.64^{+6.85}_{-1.81}\,$ & $3.74^{+0.33}_{-0.17}$ & $-0.20^{+0.14}_{-0.09}$ & $0.51^{+0.12}_{-0.07}$ & SpeX & Retrieval & - & \citet{Zalesky_2022} \\
$721.76^{+10.82}_{-11.62}$ & $4.5^{+0.07}_{-0.07}$ & $0.18^{+0.04}_{-0.04}$ & $1.97^{+0.13}_{-0.14}$ & FIRE & Retrieval & - & \citet{Gaarn_2023} \\ 
\enddata
\tablenotetext{a}{These models assume solar abundances.} 
\tablenotetext{b}{Luminosity analysis refers to using spectroscopic and photometric observations, along with atmosphere models, to estimate the bolometric luminosity of an object. Combined with age estimates, the bolometric luminosity can be used to estimate a range of $T_\mathrm{eff}$ and $\log(g)$ from evolutionary tracks, as described in \citet{Burningham_2011}.}  
\tablenotetext{c}{SM refers to atmosphere models from the Saumon \& Marley group \citep{Saumon_2006, Saumon_2008, Saumon_2012, Morley_2012}.}
\end{deluxetable*}
\end{rotatetable}

\begin{deluxetable*}{lccccccccc}
\tablenum{5}
\tablecaption{Best-fit parameters derived for GJ 570D full-spectrum and individual-band fits \label{tab:GJ_570D_fit_table}}
\tablewidth{0pt}
\tablehead{\colhead{Fitted wavelengths ($\mu$m)} & \colhead{$\chi_\nu^2$} & $\nu$ & \colhead{$T_\mathrm{eff}$ (K)} & \colhead{$\log(g)$} & \colhead{$Z$} & 
           \colhead{C/O} \\ 
    }
\startdata
$0.8-2.5$ & 26.1 & 3927 & $820^{+50}_{-50}\,$ & $4.6^{+0.25}_{-0.25}$ & $+0.0^{+0.25}_{-0.25}$ & $0.48^{+0.23}_{-0.1}$ \\ 
$0.9-1.11$ & 3.6 & 726 & $760^{+50}_{-50}\,$ & $5.5^{+0.25}_{-0.25}$ & $+0.4^{+0.25}_{-0.25}$ & $0.58^{+0.23}_{-0.1}$ \\ 
$1.14-1.35$ & 30.9 & 584 &$720^{+50}_{-50}\,$ & $5.4^{+0.25}_{-0.25}$ & $+0.1^{+0.25}_{-0.25}$ & $0.65^{+0.23}_{-0.1}$ \\ 
$1.48-1.79$ & 17.8 & 657 & $760^{+50}_{-50}\,$ & $5.5^{+0.25}_{-0.25}$ & $+0.5^{+0.25}_{-0.25}$ & $0.65^{+0.23}_{-0.1}$ \\ 
$1.90-2.31$ & 4.4 & 676 & $820^{+50}_{-50}\,$ & $5.5^{+0.25}_{-0.25}$ & $+0.3^{+0.25}_{-0.25}$ & $0.58^{+0.23}_{-0.1}$\\ 
\enddata
\end{deluxetable*}

\begin{deluxetable*}{lccccccccc}
\tablenum{6}
\tablecaption{ Best-fit parameters derived for HD 3651B full-spectrum and individual-band fits \label{tab:HD_3651B_fit_table}}
\tablewidth{0pt}
\tablehead{\colhead{Fitted wavelength range ($\mu$m)} & \colhead{$\chi_\nu^2$} & $\nu$ & \colhead{$T_\mathrm{eff}$ (K)} & \colhead{$\log(g)$} & \colhead{$Z$} & 
           \colhead{C/O} \\ 
    }
\startdata
$0.8-2.5$ & 25.0 & 3923 & $740^{+50}_{-50}\,$ & $4.3^{+0.25}_{-0.25}$ & $-0.1^{+0.25}_{-0.25}$ & $0.44^{+0.23}_{-0.1}$ \\ 
$0.9-1.11$ & 3.1 & 734 & $720^{+50}_{-50}\,$ & $5.5^{+0.25}_{-0.25}$ & $+0.3^{+0.25}_{-0.25}$ & $0.52^{+0.23}_{-0.1}$ \\ 
$1.14-1.35$ & 33.9 & 591 & $740^{+50}_{-50}\,$ & $5.5^{+0.25}_{-0.25}$ & $0.3^{+0.25}_{-0.25}$ & $1.0^{+0.23}_{-0.1}$ \\ 
$1.48-1.79$ & 13.9 & 664 & $720^{+50}_{-50}\,$ & $5.5^{+0.25}_{-0.25}$ & $+0.5^{+0.25}_{-0.25}$ & $0.58^{+0.23}_{-0.1}$ \\ 
$1.90-2.31$ & 3.0 & 683 & $780^{+50}_{-50}\,$ & $5.3^{+0.25}_{-0.25}$ & $+0.3^{+0.25}_{-0.25}$ & $0.58^{+0.23}_{-0.1}$\\ 
\enddata
\end{deluxetable*}

\begin{deluxetable*}{lccccccccc}
\tablenum{7}
\tablecaption{Best-fit parameters derived for Ross 458C full-spectrum and individual-band fits \label{tab:Ross_458C_fit_table}}
\tablewidth{0pt}
\tablehead{\colhead{Fitted wavelength range ($\mu$m)} & \colhead{$\chi_\nu^2$} & $\nu$ & \colhead{$T_\mathrm{eff}$ (K)} & \colhead{$\log(g)$} & \colhead{$Z$} & 
           \colhead{C/O} \\ 
    }
\startdata
$0.8-2.5$ & 11.7 & 3907 & $740^{+50}_{-50}\,$ & $3.9^{+0.25}_{-0.25}$ & $+0.1^{+0.25}_{-0.25}$ & $0.48^{+0.23}_{-0.1}$ \\ 
$0.9-1.11$ & 2.5 & 728 & $700^{+50}_{-50}\,$ & $5.2^{+0.25}_{-0.25}$ & $+0.5^{+0.25}_{-0.25}$ & $0.58^{+0.23}_{-0.1}$ \\ 
$1.14-1.35$ & 9.2 & 587 & $720^{+50}_{-50}\,$ & $3.1^{+0.25}_{-0.25}$ & $-0.2^{+0.25}_{-0.25}$ & $0.41^{+0.23}_{-0.1}$ \\ 
$1.48-1.79$ & 5.6 & 661 & $720^{+50}_{-50}\,$ & $5.4^{+0.25}_{-0.25}$ & $+0.5^{+0.25}_{-0.25}$ & $0.73^{+0.23}_{-0.1}$ \\ 
$1.90-2.31$ & 3.2 & 679 & $860^{+50}_{-50}\,$ & $5.0^{+0.25}_{-0.25}$ & $+0.5^{+0.25}_{-0.25}$ & $0.65^{+0.23}_{-0.1}$\\ 
\enddata
\end{deluxetable*}

%\begin{figure}[ht!]
%\plotone{figures/HD3651B_model_fit_fullspec.pdf}
%\caption{The best-fit \texttt{ATMO} spectrum (\textit{orange}) to the GNIRS spectrum of HD 3651B (\textit{black}). The uncertainties on the GNIRS spectrum are plotted as the blue line. \label{fig:HD3651B_model_fit_fullspec}}
%\end{figure}

%\begin{figure}[ht!]
%\plotone{figures/ROSS_458C_model_fit_fullspec.pdf}
%\caption{The best-fit \texttt{ATMO} spectrum (\textit{orange}) to the GNIRS spectrum of Ross 458C (\textit{black}). The uncertainties on the GNIRS spectrum are plotted as the blue line. \label{fig:ROSS_458C_model_fit_fullspec}}
%\end{figure}

%\begin{figure}[ht!]
%\plotone{figures/SpeX_Ross_458C_model_fit_fullspec.pdf}
%\caption{The best-fit \texttt{ATMO} spectrum (\textit{orange}) to the IRTF/SpeX spectrum of Ross 458C %(\textit{black}) \citep{Zhang_2021_a}. The uncertainties on the SpeX spectrum are plotted as the blue line. %\label{fig:SpeX_Ross458C_model_fit}}
%\end{figure}

\newpage
%% For this sample we use BibTeX plus aasjournals.bst to generate the
%% the bibliography. The sample631.bib file was populated from ADS. To
%% get the citations to show in the compiled file do the following:
%%
%% pdflatex sample631.tex
%% bibtext sample631
%% pdflatex sample631.tex
%% pdflatex sample631.tex

%% This command is needed to show the entire author+affiliation list when
%% the collaboration and author truncation commands are used.  It has to
%% go at the end of the manuscript.
%\allauthors

%% Include this line if you are using the \added, \replaced, \deleted
%% commands to see a summary list of all changes at the end of the article.
%\listofchanges

\end{document}